\pdfoutput=1

\documentclass[aps,prd,twocolumn,superscriptaddress,tightenlines,nofootinbib]{revtex4-1}

\usepackage{amsfonts,amsmath,amsthm,amssymb,txfonts}
\usepackage{mathrsfs}
\usepackage{bm}
\usepackage{color}
\usepackage{epsfig}
\usepackage{mathrsfs}

\usepackage[usenames,dvipsnames]{xcolor}
\definecolor{orange}{cmyk}{0,0.5,1,0}
\definecolor{rossoCP3}{cmyk}{0,.88,.77,.40}
\definecolor{graa}{rgb}{0.8,0.8,0.8}
\definecolor{blaa}{rgb}{0.2,0.2,0.6}

\newcommand{\postscript}[2]{\setlength{\epsfxsize}{#2\hsize}
   \centerline{\epsfbox{#1}}}
\newcommand{\comment}[1]{}

\begin{document}

\title{\color{rossoCP3}{Ultrahigh energy cosmic ray nuclei from
    remnants of dead quasars}}

\author{Roberto J. Moncada}
\affiliation{Department of Astrophysics, American Museum of Natural  History, Central Park West  79 St., NY 10024, USA}

\affiliation{Department of Physics, City College, City University of
  New York, NY 10031, USA
}

\author{Rafael A. Colon}
\affiliation{Department of Astrophysics, American Museum of Natural History, Central Park West  79 St., NY 10024, USA}
\affiliation{Department of Physics and Astronomy, Lehman College, City University of
  New York, NY 10468, USA
}

\author{Juan J. Guerra}

\affiliation{Department of Astrophysics, American Museum of Natural History, Central Park West  79 St., NY 10024, USA}
\affiliation{Department of Physics and Astronomy, Lehman College, City University of
  New York, NY 10468, USA
}

\author{Matthew J. O'Dowd}

\affiliation{Department of Astrophysics, American Museum of Natural History, Central Park West  79 St., NY 10024, USA}

\affiliation{Department of Physics and Astronomy, Lehman College, City University of
  New York, NY 10468, USA
}

\affiliation{Department of Physics, Graduate Center, City University
  of New York, 365 Fifth Avenue, NY 10016, USA
}

\author{Luis A. Anchordoqui}

\affiliation{Department of Astrophysics, American Museum of Natural History, Central Park West  79 St., NY 10024, USA}

\affiliation{Department of Physics and Astronomy, Lehman College, City University of
  New York, NY 10468, USA
}

\affiliation{Department of Physics, Graduate Center, City University
  of New York, 365 Fifth Avenue, NY 10016, USA
}

\begin{abstract}
  \noindent We re-examine the possibility of ultrahigh energy cosmic
  rays being accelerated in nearby dormant quasars. We particularize
  our study to heavy nuclei to accommodate the spectrum and nuclear
  composition recently reported by the Pierre Auger
  Collaboration. Particle acceleration is driven by the
  Blandford-Znajek mechanism, which wires the dormant spinning black
  holes as Faraday unipolar dynamos. We demonstrate that energy losses
  are dominated by photonuclear interactions on the ambient photon
  fields. We argue that the local dark fossils of the past quasar
  activity can be classified on the basis of how source parameters
  (mass of the central engine and photon background surrounding the
  accelerator) impact the photonuclear interaction.  In this
  classification it is possible to distinguish two unequivocal type of
  sources: those in which nuclei are completely photodisintegrated
  before escaping the acceleration region and those in which photopion
  production is the major energy damping mechanism.  We further argue
  that the secondary nucleons from the photodisintegrated nuclei (which
  have a steep spectral index at injection) can populate the energy region below
  ``the ankle'' feature in the cosmic ray spectrum, whereas heavy and
  medium mass nuclei (with a harder spectral index) populate the
  energy region beyond ``the ankle'', all the way to the high energy
  end of the spectrum. In addition, we show that five potential quasar
  remnants from our cosmic backyard correlate with the hot-spot
  observed by the Telescope Array.
\end{abstract}
\date{January 2017}
\pacs{}

\maketitle
\section{Introduction}

The origin(s) of ultrahigh energy cosmic rays
(UHECRs) remains a central question in high energy
astrophysics~\cite{Torres:2004hk,Kotera:2011cp}. 
Three principle observables drive the search for and characterization of cosmic ray sources: the energy spectrum,
the nuclear composition, and the distribution of arrival directions.
The  cosmic ray energy spectrum encompasses a plummeting flux that  drops from 
$10^4~{\rm m}^{-2} \, {\rm s}^{-1}$ at $E \sim 1~{\rm GeV}$ to
$10^{-2}~{\rm km}^{-2} \, {\rm yr}^{-1}$ at $E \sim 10^{11}~{\rm
  GeV}$. Its shape is remarkably featureless and can be described by a
broken power law: the spectrum steepens gradually from $E^{-2.7}$ to
$E^{-3.3}$, then flattens to $E^{-2.6}$ near $10^{9.6}~{\rm GeV}$
forming a feature known as ``the ankle'', and drops sharply at around
$10^{10.5}~{\rm
  GeV}$~\cite{Abbasi:2007sv,Abraham:2008ru,Abraham:2010mj}. The small
variations of the spectral index can be interpreted either as a
transition between cosmic ray populations, or as an imprint of cosmic ray propagation
effects.

The simplest interpretation of the ankle is that above $10^{9.6}~{\rm
  GeV}$ a new population emerges which dominates the more steeply
falling Galactic population of heavy nuclei. The extragalactic
component can be dominated either by protons~\cite{Bird:1993yi} or
heavies~\cite{Allard:2005ha,Allard:2005cx}. Proton-dominance beyond
the ankle is ultimately limited by the onset of photopion production
on the cosmic microwave background, whereas dominance of a heavy
composition is restricted by nucleus photodisintegration through the
giant dipole resonance  --
the so called {\it Greisen-Zatsepin-Kuz'min} (GZK) suppression at
around $10^{10.5}~{\rm GeV}$~\cite{Greisen:1966jv,Zatsepin:1966jv}.  It has also been advocated that
the ankle feature could be well reproduced by a proton-dominated power-law
spectrum, where the ankle is formed as a {\it dip} in the spectrum
from the energy loss of protons via Bethe-Heitler pair
production~\cite{Hillas:1967,Berezinsky:2002nc}. In this case
extra-galactic protons would already have started to dominate the
spectrum somewhat beyond $10^{8.7}~{\rm GeV}$.

Optical observations of air showers with fluorescence telescopes or
non-imaging Cherenkov detectors consistently find a predominantly
light composition at around $10^9~{\rm GeV}$~\cite{Kampert:2012mx} and
the contribution of protons to the overall cosmic ray flux is
$\agt$~50\% in this energy range~\cite{Aab:2014kda,Aab:2014aea,Aab:2016htd,Aab:2016zth}.  Due
the absence of a large anisotropy in the arrival direction of cosmic
rays below the ankle~\cite{Auger:2012an,ThePierreAuger:2014nja}, we
can conclude that these protons must be of extra-galactic origin.  At
energies above $10^{10}~{\rm GeV}$, the high-statistics data from the
Pierre Auger Observatory suggests a gradual increase of the fraction
of heavy nuclei in the cosmic ray flux~\cite{Aab:2014aea,
  Aab:2014kda,Aab:2016htd,Aab:2016zth}. Within uncertainties, the data
from the Telescope Array (TA) is consistent with these
findings~\cite{Abbasi:2014sfa, Abbasi:2015xga}. Moreover, the
Telescope Array has observed a statistically significant excess in
cosmic rays with energies above $10^{10.7}~{\rm GeV}$ in a region of
approximately 1150 square degrees centered on equatorial coordinates
(R.A. = $146.7^\circ$, Dec. = $43.2^\circ$) -- the TA hot spot~\cite{Abbasi:2014lda}. The
absence of a concentration of nearby sources in this region of the sky
favors a heavy nucleus hypothesis, whereby a few local sources within
the GZK sphere can
produce the hot spot through deflection in the Galactic ${\bm
  B}$-field. All in all, the apparent dominance of heavy nuclei in the
vicinity of $10^{10}~{\rm GeV}$ strongly argues against the
interpretations of the ankle given above.

One can (of course) accommodate the data with the addition of an {\it
  ad hoc} light extragalactic component below the ankle, with a steep
injection spectrum~\cite{Aloisio:2013hya}. However, a more natural
explanation of the entire spectrum and composition emerges while
accounting for the ``post-processing'' of UHECRs through
photodisintegration in the environment surrounding the
source~\cite{Unger:2015laa,Farrar:2015ikt,Globus:2015xga,Anchordoqui:2014pca}.  In such 
models relativistic nuclei accelerated by a central engine to
extremely high energies remain trapped in the turbulent magnetic field
of the source environment. Their escape time decreases faster than the
interaction time with increasing energy, so that only the highest
energy nuclei can escape the source unscathed. In effect, the
source environment acts as a high-pass filter on the spectrum of
cosmic rays. All nuclei below the energy filter interact, scattering
off the far-infrared photons in the source environment.  These
photonuclear interactions produce a steep spectrum of secondary
nucleons, which is overtaken by the harder spectrum of the surviving
nucleus fragments above about $10^{9.6}~{\rm GeV}$. These overlapping
spectra could then carve an ankle-like feature into the source
emission spectrum. The spectrum above the ankle exhibits a progressive
transition to heavy nuclei, as the escape of non-interacting nuclei
becomes efficient. In such models, the source evolution with
redshift plays a paramount role on the source parameters. For models
with positive redshift evolution (a greater number of sources per
comoving volume at high redshift), both the imposition of a dominant
light extragalactic component below the ankle and the source-ankle
model favor a hard injection spectrum $\propto E^{-1}$. However, it
has been recently noted that fitting the high end of the cosmic ray (CR) spectrum
and composition with negative source evolution allows for softer
injection spectra~\cite{Taylor:2015rla}. 

On a separate track, it was recently pointed out that Fermi-LAT
observations~\cite{Ackermann:2014usa,TheFermi-LAT:2015ykq} severely
constrain the distribution of UHECR sources~\cite{Liu:2016brs}.  This
is because on the way to Earth UHECRs interact with the radiation
fields permeating the universe and give rise to energetic
electron/positron pairs and photons, which in turn feed
electromagnetic cascades resulting in a diffuse gamma-ray background
radiation. The recent study in~\cite{Liu:2016brs} suggests that a
local ``fog'' of UHECRs accelerated in nearby sources must exist so as
to not overproduce the cascade gamma-ray flux beyond Fermi-LAT
observations.

In consideration of these new results, it seems interesting to explore
whether remnants of dead quasars, which are distributed preferentially
in the low-redshift universe, could accelerate heavy nuclei up to the
highest observed energies.  Actually, nearby quasar remnants have long
been suspected to be sources of
UHECRs~\cite{Boldt:1999ge,Boldt:2000dx,Torres:2002bb,Neronov:2007mh}.  Particle
acceleration proceeds via the Blandford-Znajek mechanism, which wires
 dormant spinning black holes as  Faraday unipolar
dynamos~\cite{Blandford:1977ds,Znajek}. Detailed modeling with
reasonable assumptions on source parameters suggests that protons can
be efficiently launched at $E \agt 10^{12}~{\rm GeV}$.  Photopion
production in collisions with ambient photons becomes a relatively
important effect during the final phase of the acceleration process,
and could damp the maximum attainable energy to values well below that
imposed by the voltage drop. The large charge $Ze$ of heavy nuclei
facilitates their acceleration to highest energy (scaling linearly
with $Z$). However, heavy nuclei could also interact with the ambient
photons and may suffer significant spallation. In this paper we
study in detail the dominant processes for nucleus energy losses,
while searching for plausible source parameters which may allow
acceleration of heavy nuclei up to the highest observed energies.

The layout of the paper is as follows. In Sec.~\ref{sec:2} we provide
an assessment of what is known observationally about the evolution of
quasar activity and then discuss the general framework use to relate
the quasar population with massive dark objects at the center of local
inactive galaxies. We also show that five nearby relics of such past
activity correlate with the TA hot-spot. In Sec.~\ref{sec:3} we review
the generalities of particle acceleration near the event horizon of a
spinning supermassive black hole and determine plausible source
conditions for acceleration of UHECR nuclei. In Sec.~\ref{sec:4} we
examine the various energy-loss mechanisms of one-shot acceleration,
including radiative processes and photonuclear interactions with the
ambient photons. We show that, in general, photonuclear energy losses
are dominant. After that we propose a novel  phenomenological model to
explain the evolution of the UHECR spectrum and composition
with energy, without fine-tuning. Finally, in Sec.~\ref{sec:5} we
present our conclusions.

\section{Dormant black holes after a shining past}
\label{sec:2}

The first extensive radio surveys of the 1950s revealed a new family
of radio sources that, by the early 60s, had been associated with
point-like optical counterparts~\cite{Hazard}. These quasi-stellar
radio sources, or quasars, 
revealed very large systemic redshifts indicating cosmological
distances~\cite{Schmidt}. This implied luminosities up to hundreds of
times that of the Milky Way, apparently emitted from very small
size-scales~\cite{Smith}, suggesting an extremely energetic emission
mechanism. It has now been established that quasars are ultimately
powered by the accretion of matter onto the $10^7-10^{10} M_\odot$
supermassive black holes (SMBHs) at cores of
galaxies~\cite{Zeldovich,Salpeter:1964kb,LyndenBell:1969yx,Antonucci:1993sg,Urry:1995mg}.

Quasars and their relativistically-beamed counterparts, blazars, are
the most luminous sub-class of {\it active galactic nuclei} (AGNs),
which is the general name for an actively-accreting SMBH. In fact,
these objects are the most luminous continuously-emitting objects in
the Universe. Quasar luminosity is, to first order, constrained by the
Eddington limit~\cite{Eddington}. This is the maximum luminosity of a spherically
accreting object, determined by the point at which outward radiation
pressure halts gravitational infall. The Eddington limit is
proportional to the mass of the accreting object. Scaled for a quasar
powered by a SMBH of $10^9~M_\odot$, this luminosity is:
\begin{equation}
L_{\rm Edd} = 1.3 \times 10^{47} \, \frac{M}{10^9 M_\odot}~{\rm erg/s} \sim 10^2 L_{\rm galaxy} \,,
\end{equation}
where $M_\odot$ is the solar mass and $L_{\rm galaxy}$ is the
bolometric luminosity of large galaxies like the Milky Way. Note that
accretion by the standard $\alpha$-disk, which is typically assumed
for quasars, is also Eddington limited~\cite{Shakura:1972te}.

\begin{figure}[b]
\postscript{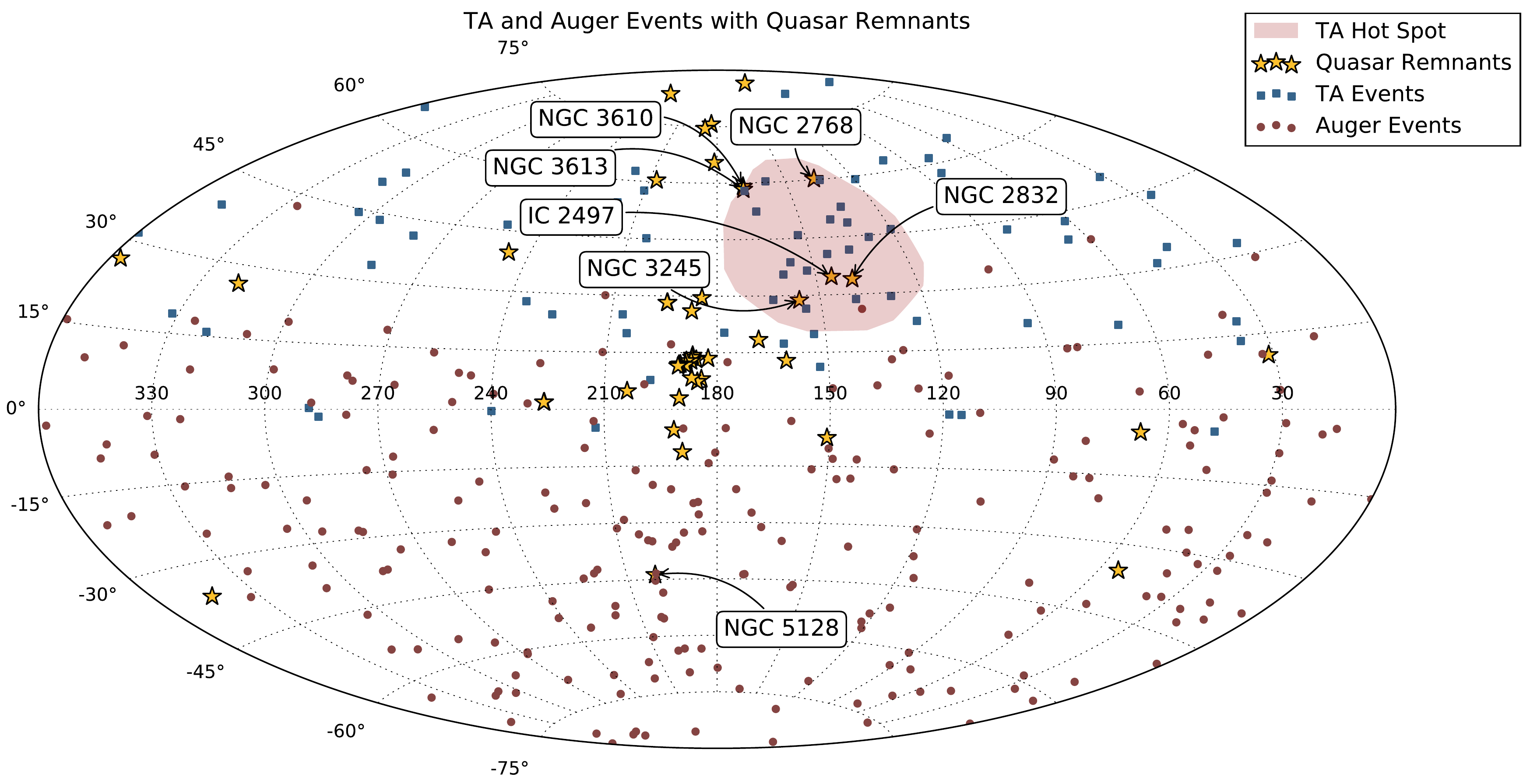}{0.99}
\caption{Comparison of UHECR event locations with quasar remnants in
  equatorial coordinates, with R.A. increasing from right to left.  The
  circles indicate the arrival directions of 231 events with $E >
  52~{\rm EeV}$ and zenith angle $\theta < 80^\circ$ detected by the
  Pierre Auger Observatory from 2004 January 1 up to 2014 March
  31~\cite{PierreAuger:2014yba,Abreu:2010ab,Abraham:2007si,Abraham:2007bb}.
  The squares indicate the arrival directions of 72 events with
  $E>57~{\rm EeV}$ and $\theta < 55^\circ$ recorded from 2008 May 11
  to 2013 May 4 with TA~\cite{Abbasi:2014lda}. The stars indicate the
  location of nearby quasar
  remnants~\cite{Torres:2002bb,Isola:2003jk}.  The shaded region
  delimits the TA hot-spot. The locations of the sources inside the
  TA hot-spot and the closest object in the survey (NGC 5128) are
  explicitely identified.}
\label{fig:1}
\end{figure}

It is generally accepted that the number density of quasars peaked at
redshift $z \sim 2$, when the universe was one-fifth of its present
age~\cite{Rees}. Locally, essentially all galaxies with spheroidal
components contain a SMBH~\cite{Richstone:1998ky,Ho:1998hz,Kormendy:2004tc}.  The close
relationships between the masses of these SMBHs and their host
galaxy's properties, including stellar velocity dispersion, spheroidal
luminosity, and dynamical mass, indicates that the growth of SMBHs and
their hosts are closely
tied~\cite{Kormendy:1995er,Magorrian:1997hw,Gebhardt:2000fk,Ferrarese:2000se,Haring:2004hr}. These
relationships also allow us to make a relatively direct determination
of the local SMBH mass
function~\cite{Vika:2009ef,McLure:2003em,Graham:2007qb,Marconi:2003tg,Tamura:2005kc}.

Another approach to deriving the local SMBH mass function is to
integrate SMBH accretion over cosmic time based on the evolving
high-$z$ quasar luminosity function
\cite{Marconi:2003tg,Tamura:2005kc,Soltan:1982vf,Chokshi,
  Small92,Cao:2008pd,Li:2011ik,Shankar:2007zg}. Such models are able
to reproduce the ``observered'' low-$z$ mass function for quiescent
SMBHs very successfully by assuming the typical radiative efficiencies
for an $\alpha$-disk ($0.1\alt \eta \alt 0.2$), and reasonable
Eddington fractions --- quasar luminosities as a fraction of the
Eddington limit (Eddington fraction) --- of $0.2 \alt L/L_{Edd}\alt
1$~\cite{Graham:2007qb,Marconi:2003tg,
  Cao:2008pd,Li:2011ik,Yu:2002sq}. This general agreement supports for
the idea that accretion in AGN mode is an important, and perhaps the
principle mechanism by which local SMBHs grow~\cite{Rees,Small92,Marconi:2003tg,Cao:2008pd}. Thus, it is reasonable
to assume that many of the more massive quiescent SMBHs in the local
universe are "remnant'' quasars~\cite{Rees,Chokshi}, and that these greatly outnumber
active quasars in the modern epoch.

In their active phases, accreting SMBHs are expected to support strong
magnetic fields. Assuming a rapidly-rotating black hole, which may be
common in luminous quasars~\cite{Cao:2008pd,Volonteri:2007tu}, this
magnetic field is expected to survive the expiration of the quasar
phase. We calculate a typical relic magnetic field strength for a
rapidly rotating SMBH by assuming that it corresponds to maximum field
strength acquired during its quasar phase.  This in
turn corresponds to its maximum accretion rate. To determine this
accretion rate we assume the lower end of $\alpha$-disk radiative
efficiency, $\eta = 0.1$ (corresponding to a higher accretion rate per
unit luminosity), and that the quasar radiates at the Eddington
limit. The mass accretion rate needed to sustain the Eddington
luminosity is
\begin{equation}
\dot{M}_{\rm Edd} = \frac{L_{\rm Edd}}{\eta c^2} \simeq 22  \ M_9 \ M_\odot~{\rm yr}^{-1} \, .
\end{equation}
where $M_9 =M/(10^9 M_\odot)$. The accreting plasma is assumed to
support an axisymmetric magnetic field configuration due to the
generation of currents. Following~\cite{Boldt:1999ge}, we get an
estimate of the characteristic magnetic field strength $B_0$ by
assuming pressure equilibrium between the magnetic field and the
in-falling matter~\cite{Anchordoqui:2016dcp}
\begin{equation}
B_0 \sim 10^4 \ M_9^{-1/2} \ \left(\frac{\dot M}{\dot M_{\rm Edd}}
\right)^{1/2}~{\rm G} \, .
\label{Bp}
\end{equation}
This magnetic field may be taken as a fiducial strength for quasars
remnants at the upper end of the local SMBH mass and spin functions. 

This represents the magnetic field strength developed by a maximally
rotating black hole at the peak of its accretion phase and at the
upper end of the mass function. In the following calculations we work
with this fiducial strength, yielding. It will be seen that a relatively high
field strength, $B_0 \sim 10^4~{\rm G}$, is necessary to achieve a significant flux of
$10^{11}~{\rm GeV}$ single-proton CRs around a $10^9 \, M_\odot$
SMBH. However, when we consider black holes up to $10^{10} \,
M_\odot$, and heavy nuclei CRs, the required field strength drops by
up to two orders of magnitude. We also assume that a significant
fraction of the maximum field strength is sustained as accretion rate
drops into the remnant quasar phase. Further data on the spatial
association of UHECRs with SMBHs will test this assumption.

Significant magnetic fields do appear to survive in the case of least
one class of remnant quasar. BL Lacertae objects are AGNs of the
blazar class -- broadly, AGNs with jets that are closely aligned with our
line of sight~\cite{Urry:1995mg,Blandford:1978}. BL Lacs show little
or no signs of broad emission lines or thermal
emission~\cite{Urry:1995mg,Marcha,Giommi:2011sn}. This implies that at
least a subset of BL Lacs are starved of gas, and perhaps in a
post-quasar phase. Under this interpretation, flat-spectrum radio
quasars (FSRQs) become the progenitor population of low-accretion BL
Lacs~\cite{Cavaliere:2001gb,Boettcher:2001jp,Ajello:2011zi}. This evolutionary argument is supported by BL Lacs'
negative cosmological evolution; their number densities rise at around
the same epoch as the decline of FSRQs~\cite{Ajello:2013lka}. BL Lacs
exhibit BZ-powered jets~\cite{Maraschi:2002pp}, indicating that
significant magnetic fields may be sustained long after the cessation
of significant accretion.

Hanny's Voorwerp contains perhaps the first direct probe of
quasar  evolution~\cite{Lintott:2009wq}. Observations in the
optical, ultraviolet, and X-ray indicate that this unusual object near
the spiral galaxy IC 2497 contains highly ionized gas~\cite{Jozsa:2009qm}. The
emission-line properties, and lack of X-ray emission from IC 2497,
seem to indicate this spiral galaxy is a highly obscured AGN with a
novel geometry arranged to allow photoionization of Hanny's Voorwerp
but not the galaxy's own circumnuclear gas~\cite{Rampadarath:2010yv}, or alternatively the
luminosity of the central source has decreased dramatically within the
last $10^5$~yr~\cite{Schawinski:2010up}. Should this be the case, Hanny's Voorwerp may
exemplify the first detection of a quasar light echo. Interestingly,
IC 2497 is inside the TA hot spot; see Fig.~\ref{fig:1}. 
However, the measured redshift $z =0.05$ of IC 2497, sets this dormant quasar outside our local GZK--horizon ($\sim
100$~Mpc). To sample sources inside the GZK--sphere we adopt a
distance-limited ($z \alt
0.02$) compilation of quasar remnant candidates, capable of
accelerating UHECRs~\cite{Torres:2002bb,Isola:2003jk}. It is compelling that five of
the objects contained in this sample are also inside the TA hot spot; see Fig.~\ref{fig:1}.

Centaurus A (Cen A) is a complex radio-loud source
  identified at optical frequencies with the galaxy NGC 5128. Cen A is
  the closest radiogalaxy to Earth (distance $\sim 3.4~{\rm Mpc}$) and has long
been suspected to be a potential UHECR
accelerator~\cite{Cavallo:1978,Romero:1995tn}. The Pierre Auger
Collaboration has searched for anisotropies in the direction of Cen A
scanning the energy threshold between $10^{10.6}~{\rm GeV}$ and
$10^{10.9}~{\rm GeV}$ and counting events in angular radii ranging
from $1^\circ$ to $30^\circ$~\cite{PierreAuger:2014yba}.  The
strongest departure from isotropy (post-trial probability $\sim
1.4\%$) has been observed for $E > 58~{\rm EeV}$ in a window of
$15^\circ$: 14 events (out of a total of 155) have been observed in
such an angular window while 4.5 are expected on average from
isotropic distributions.

 A statistical analysis to ascertain the significance of a
  possible correlation between UHECRs and SMBHs would require using
  data from Auger and TA. However, if the high energy end of the
  cosmic ray spectrum is the same in the north and in the south, there
  appears to be a difference in the energy scale between the two
  detectors~\cite{Dawson:2013wsa}, making such study unreliable at present.  Future
  experiments that will observe both hemispheres using the same
  apparatus, such as the Probe Of Extreme Multi-Messenger
  Astrophysics (POEMMA),  would make a robust
  statistical analysis more straightforward.\footnote{POEMMA has been
    recently selected by NASA for an in-depth probe mission concept
    study in preparation for the next deceadal survey~\cite{Adams:2017fjh}.}

 For convenience, we subdivide the SMBH sample into two
  subcomponents: one with masses $\agt 10^{10} M_\odot$ and one with
  masses $\alt 10^{10} M_\odot$.  The rationale for using $10^{10}
  M_\odot$ as the demarcation mass will become evident in
  Sec.~\ref{sec:4}. Roughly speaking, 16\% of SMBHs have mass above
  $10^{10} M_\odot$~\cite{Isola:2003jk}. Inside the TA hot spot one of six (five nearby) candidate
  sources is consistent with a mass$ \agt 10^{10} M_\odot$. Thus, we
  can assume that the ratios inside the hot spot are representative of
  the ratios of the universe at large.

\section{UHECR acceleration in quasar remnants}
\label{sec:3}

In realistic astrophysical situations involving quasars, the black hole
itself is uncharged, and the gravity of the accretion disk is practically
negligible. This means that the geometry of spacetime  is described by the
Kerr metric and therefore determined by two free parameters: the black hole
mass $M$ and its
angular momentum 
\begin{equation}
|\bm{J}| = a \, J_{\rm max} = a \, \frac{GM^2}{c} \,,
\end{equation}
where $0 \leq a \leq 1$ is the dimensionless spin parameter~\cite{Kerr:1963ud}.  Before
proceeding we note  that for a Schwarzschild
black hole, $a = 0$~\cite{Schwarzschild:1916uq}, whereas for a maximally spinning (a.k.a. extreme Kerr)
black hole, $a =1$. As a matter of fact, the increase of the spin parameter would
stop at $a \approx 0.998$ because photons emitted from
the disk on retrograde paths are more likely to be captured by the
hole than prograde ones, hence de-spinning the hole~\cite{Thorne:1974ve}.

For static black holes, the scale characterizing
the event horizon $r_h$ is the Schwarzschild
radius
\begin{equation}
r_S = \frac{2GM}{c^2} \simeq 3 \times 10^{14} \left(\frac{M}{10^9
    M_\odot} \right)~{\rm cm} \, .
\end{equation}
For spinning black holes, the (spherical) event horizon surface is
defined by
\begin{equation}
r_h = r_g \left(1 + \sqrt{1-a^2} \right) \,,
\end{equation}
where $r_g = r_S/2$ is the gravitational radius. The
ergosphere is the region described by the relation 
\begin{equation}
r_h < r < r_{\rm max} = r_g \left(1 + \sqrt{1-a^2 \cos^2 \theta} \right) \,,
\end{equation}
where $\theta$ is the angle to the polar axis; see
e.g.~\cite{Visser:2007fj,Rieger:2011ch}. Inside the ergosphere (which
is ellipsoidal in shape), spacetime is pulled heavily towards the
direction of the black hole rotation (frame dragging), and
consequently no static observer can exist because the particles must
co-rotate with the hole. The black hole's angular velocity
$\bm{\Omega}$ coincides with the angular velocity of the dragging of
inertial frames at the horizon~\cite{Rieger:2011ch}
\begin{equation}
|\bm{\Omega}| = a \left(\frac{c}{2 r_h} \right) \, .
\end{equation}

It has long been known that a rotating black hole can radiate away its
available reducible energy~\cite{Penrose:1969pc,Penrose:1971uk}. In
this context Blandford and Znajek (BZ) proposed a model of electromagnetic
extraction of black hole's rotational energy based on the analogy with
the classical Faraday  (unipolar induction) dynamo
phenomenon~\cite{Blandford:1977ds,Znajek}. In the BZ mechanism, the
magnetic field $\bm{B}$ near the horizon taps the rotational energy of the
black hole and generates powerful outflows of electromagnetic
(Poynting) energy. BZ argue that spacetime frame dragging induces an
electric field $\bm{E}$ that is strong enough to break the vacuum and
establish an electron-positron force-free magnetosphere.

The particulars of the black hole magnetosphere could be very
complicated, distinctively if the poloidal magnetic field near the
horizon is misaligned with the black hole rotation axis~\cite{Neronov:2007mh}.  For
simplicity, herein we assume that the $\bm{B}$-field is aligned with the
axis of rotation. This manageable system allows a transparent
analytic calculation which captures the essence of most aspects of the
BZ mechanism and leads to a correct order of magnitude.  

Conduction electrons within the magnetosphere undergo collisions with
the orbiting atoms, and this causes them to take up the rotational
motion. At each point the electrons have a net drift velocity, ${\bf
  v} = \bm{\Omega} \wedge \bm{r}$, where $\bm{r}$ is the position
vector of the point in question relative to an origin which is chosen
to lie on the magnetic axis. The conduction electrons experience a
magnetic force $-e {\bf v} \wedge \bm{B}$, which is mainly directed
towards the central axis; as a result a negative charge appears in the
core of the magnet, and a positive charge on its curved outer
surface. In equilibrium an electrostatic field is set up, such that
the total Lorentz force on the conduction electrons is zero. Namely,
$\bm{F} = - e \bm{ E} - e {\bf v} \wedge \bm{ B} = \bm{ 0}$, and so
\begin{equation}
\bm{ E }= - {\bf v} \wedge \bm{ B} \, . 
\label{E}
\end{equation}
We now calculate the potential difference $V$ corresponding to this
electric field. It is straightforward to see by inspection of
(\ref{E}) that along the axis of rotation $\bm{E} = 0$, and therefore $V =
0$ at all points along this axis. For a maximally spinning
hole, the potential drop between the central ($r = 0$) and the
marginal ($r = r_{\rm max}$) magnetic surfaces passing through the
horizon is found to be
\begin{eqnarray}
V  & = & - \int \bm{E} \cdot d \bm{s} = \int_0^{r_{\rm max}} \Omega \, B_0 \,
r \, dr
  \sim \frac{1}{4} \, \frac{G}{c} \, MB_0  \nonumber \\
& \sim & \frac{1}{4} \frac{6.7 \times 10^{-11} }{3
  \times 10^8} \frac{1.9 \times 10^{30}}{M_\odot} M \, B_0~{\rm m^2 s^{-1}} \,,
\label{V1}
\end{eqnarray}
where $B_0$ is the magnetic field strength.
Now, recalling that $10^4~{\rm G}
=1~{\rm T}= 1~{\rm V\, s \, m^{-2}}$, we can rewrite (\ref{V1}) as
\begin{equation}
V \sim 10^{20} \, M_9  \, B_4~{\rm V} \, ,
\label{V2}
\end{equation}
where $B_4 = B_0/10^4~{\rm G}$.

\begin{figure}
\postscript{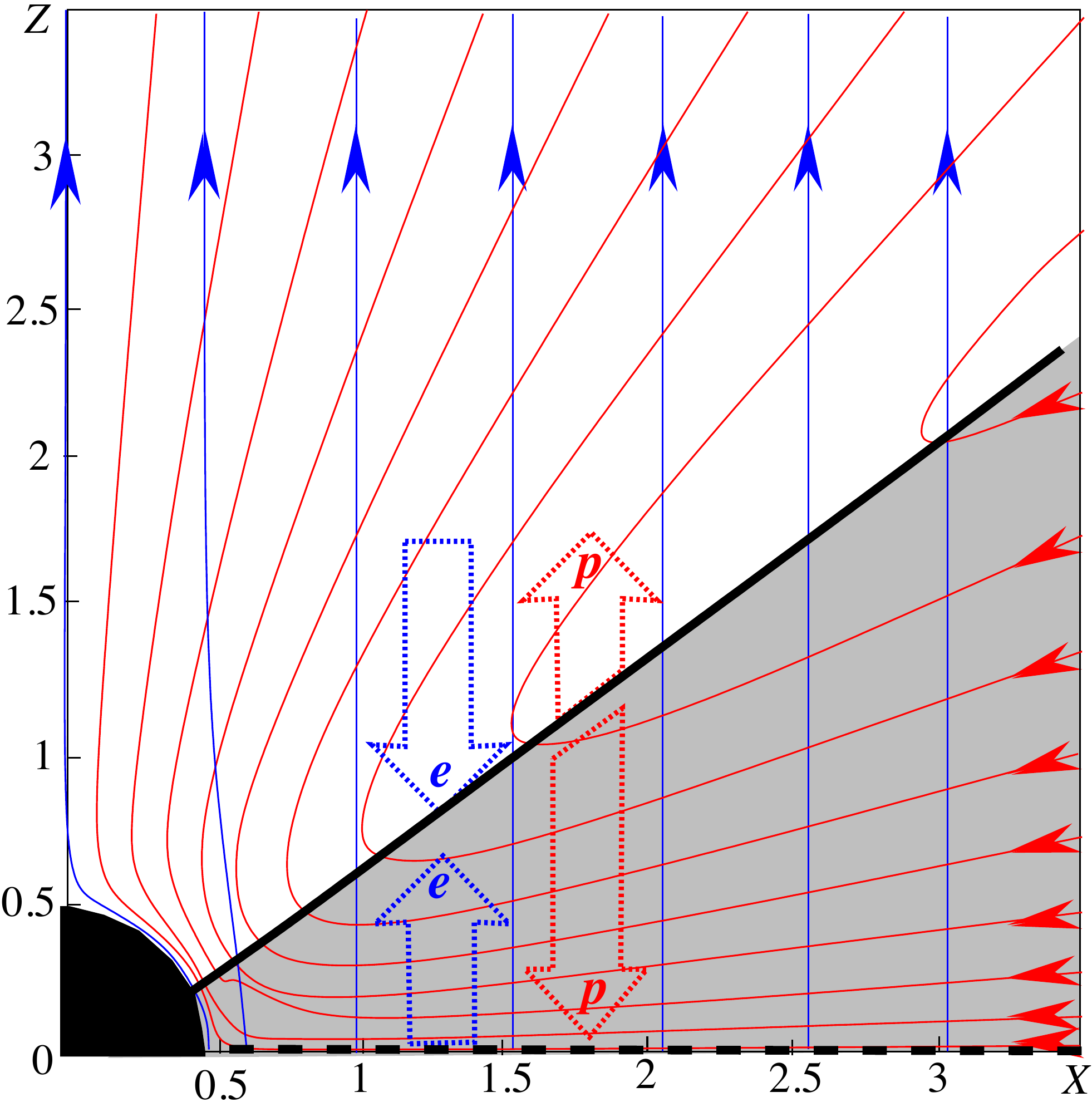}{0.9}
\caption{Magnetic (blue solid) and electric (red solid) field lines
  associated to the asymptotically homogeneous $\bm{B}$ field aligned
  with the rotation axis of a maximally spinning black hole. The wide
  blue and red arrows show the directions of acceleration of electrons
  and protons in different regions. The acceleration can be avoided
  only if a particle resides at a surface at which the electric field
  is orthogonal to the magnetic field, the so-called “force-free”
  surfaces.  Thick solid and dashed lines indicate the conical and
  equatorial force-free surfaces. The force-free surfaces separate
  different acceleration regions, with electric field directed along
  or oppositely to the magnetic field. Taken
  from~\cite{Neronov:2007mh}.}
\label{fig:field-lines}
\end{figure}

Far away from the horizon, the electromagnetic field can be expressed as~\cite{Wald:1974np}
\begin{equation}
\lim_{r \to \infty} B_{\hat r}  = B_0 \cos \theta, \quad \quad \quad
\quad \quad \quad 
\lim_{r \to \infty}  B_{\hat \theta}  = - B_0 \sin \theta \,, 
\end{equation}
and
\begin{equation}
\lim_{r \to \infty} E_{\hat r}  = - \frac{B_0 \, a M (3 \cos^2 \theta -
  1)}{r^2}, \quad 
\lim_{r \to \infty}  E_{\hat \theta}  = {\cal O} (r^{-4}) \, .
\end{equation} 
The electric field has a quadrupole topology, which is distinctly seen 
in Fig.~\ref{fig:field-lines}.  The voltage difference between the horizon and
$r = \infty$ is of order $V$. With such a huge voltage drop along
field lines, one expects that electrons roaming in the vicinity of
the horizon would be accelerated to huge energies and, upon colliding
with stray photons and/or positrons, produce a cascade of
$e^+e^-$ pairs. Very quickly therefore, the vacuum surrounding the
black hole would be filled with a highly conducting plasma. The
plasma-loaded field lines can serve as {\it wires} to complete a
circuit between the pole and equator of the horizon. The electromotive
force around this circuit is numerically comparable to
(\ref{V2}). Thus, if
a cosmic ray baryon can fully tap this potential acceleration up to
extreme energies,
\begin{equation}
E_{\rm max} = Ze V \sim 10^{11} \, Z \, M_9  \, B_4~{\rm  GeV} \sim
10^{11} \, Z\, M_9^{1/2}~{\rm GeV}\,,
\label{maxE}
\end{equation}
would become possible. However, the charge density in the vicinity of
accreting black holes could be so high that a significant fraction of
this potential would be screened and so no longer available for
particle acceleration. Therefore, it seems more appropriate to define
an effective potential where the available length scale, the gap
height $\zeta$, is explicitly taken into
account. Following~\cite{Rieger:2011ch}, we take 
\begin{equation}
V_{\rm eff} \sim V \ \left(\frac{\zeta}{r_h} \right)^2 \, .
\end{equation}
Accordingly, the characteristic rate of energy gain is found to be
\begin{equation}
\left. \frac{dE}{dt} \right|_{\rm acc} = Z e \ V_{\rm eff} \ \frac{c}{\zeta} \,,
\end{equation}
and so the acceleration timescale is given by
\begin{equation}
\tau_{\rm acc}  = E \left[\left. \frac{dE}{dt}
  \right|_{\rm acc} \right]^{-1} = \frac{E}{Z e V_{\rm eff}}  \,  \frac{\zeta}{c} \, .
\label{muddler}
\end{equation}

In a realistic situation the energy losses of the accelerated
particles limit the maximal energies to the values below the estimate
of (\ref{maxE}). It is this that we now turn to study. Before
proceeding though, it is interesting to note that the axisymmetric
stationary flow in the vicinity of the rotation-powered compact object
could provide a source of ultrarelativistic nuclei with an extremely
flat injection spectrum (typically $\propto
E^{-1}$~\cite{Blasi:2000xm,Kotera:2015pya}), in good agreement with
the requirements  to
accommodate  Auger
observations~\cite{Aab:2016zth,Unger:2015laa}. Moreover, a steepened
injection spectrum would arise quite naturally if the medium
surrounding the compact object were leaky at these late ages.

\section{Mechanisms of energy dissipation}
\label{sec:4}

\begin{table*}
\caption{Relevant source parameters and normalization factor
  $n_0^\beta$, for $\beta =  -4,\, -5$. \label{tabla}}
\begin{tabular}{ccccccc}
\hline 
\hline
Source  &~~~~~~~$L_{\rm
  FIR}/L_\odot$~~~~~~~&~~~~~~~$M_9$~~~~~~~&~~~~~~~$n_0^{-4}/({\rm MeV
  \, cm}^3)$~~~~~~~&~~~~~~~$n_0^{-5}/({\rm MeV
  \, cm}^3)$~~~~~~~&~~~~~~~$z$~~~~~~~&
References\\ 	
\hline
NGC 2768	&  $\phantom{<} 4.8 \times 10^8$ &      0.16 & $7.8 \times
10^{23}$ & $1.0 \times 10^{24}$
&  $0.0051$ & \cite{Torres:2002bb,Crocker:2008dk,Temi:2009ue}  \\
NGC 2832	& $<1.7 \times 10^8$	 &  11.4 &~~~~~~~$3.0 \times 10^{19}$~~~~~~
& ~~~~~~~$3.9 \times 10^{19}$~~~~~~~&~~~~~~~$0.0194$~~~~~~~ & \cite{Isola:2003jk,Temi:2009ue,Brown:2011xh}\\
NGC 3610 & $\phantom{<} 4.7 \times 10^7$ & 0.05 & $6.9 \times 10^{23}$ & $8.9
\times 10^{23}$ & $0.0066$  &  \cite{Torres:2002bb,Temi:2009ue,Temi:2003sj} \\ 
NGC 3613 & $< 3.6 \times 10^8$ & 0.16 & $4.8 \times 10^{23}$ & $6.2
\times 10^{23}$ & $0.0066$ & \cite{Torres:2002bb,Brown:2011xh,Temi:2003sj} \\
NGC 4125 & $\phantom{<} 3.0 \times 10^8$ & 0.28 & $1.5 \times 10^{23}$ & $1.9
\times 10^{23}$ & 0.0063& \cite{Torres:2002bb,Temi:2009ue,Temi:2003sj} \\ 
\hline
\hline
\end{tabular}
\end{table*}

\begin{figure*}[tpb] 
\begin{minipage}[t]{0.49\textwidth}
    \postscript{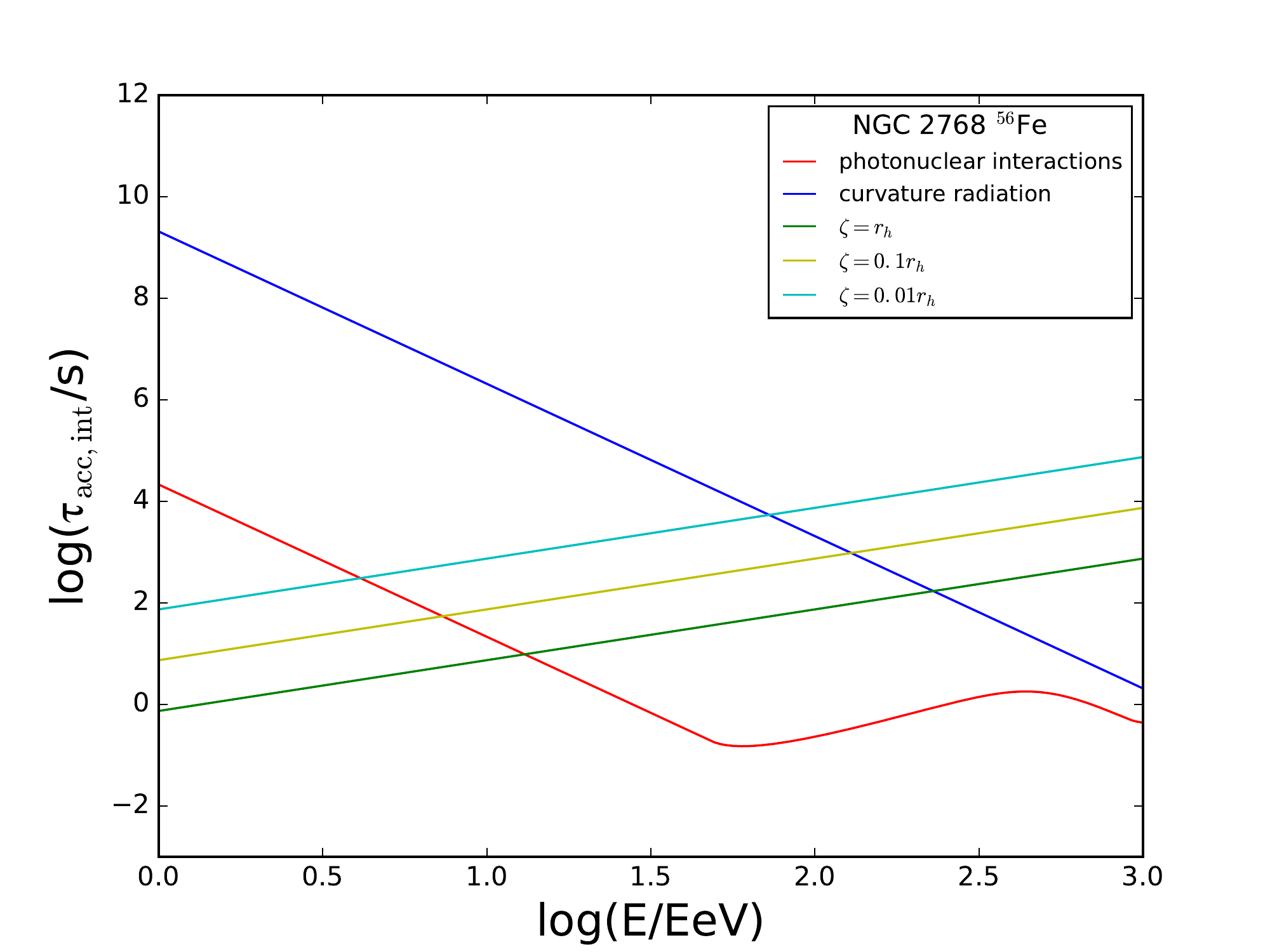}{0.99} 
\end{minipage} 
\begin{minipage}[t]{0.49\textwidth}
    \postscript{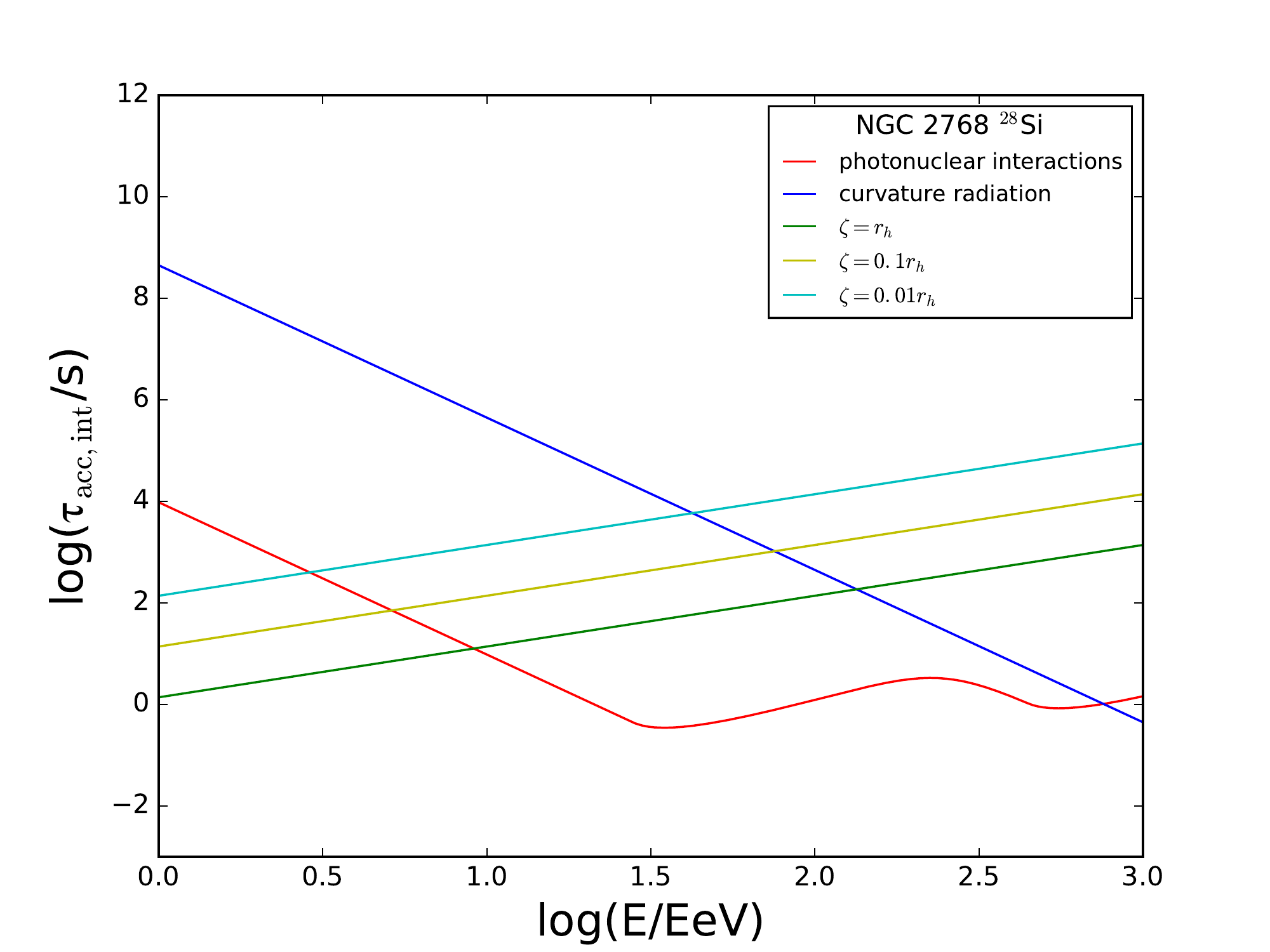}{0.99} 
\end{minipage}
\hfill \begin{minipage}[t]{0.49\textwidth}
  \postscript{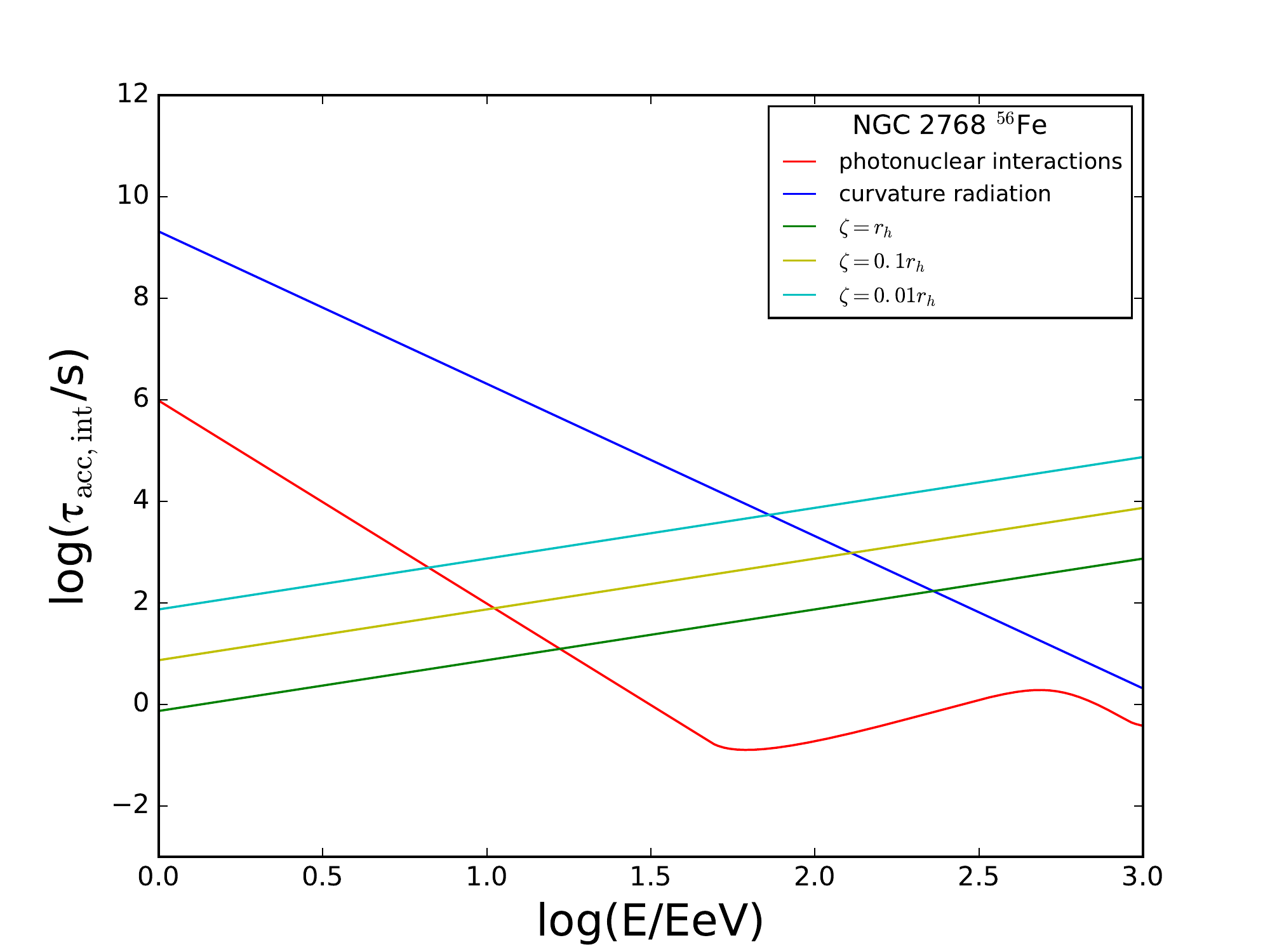}{0.99} 
\end{minipage} 
\begin{minipage}[t]{0.49\textwidth}
  \postscript{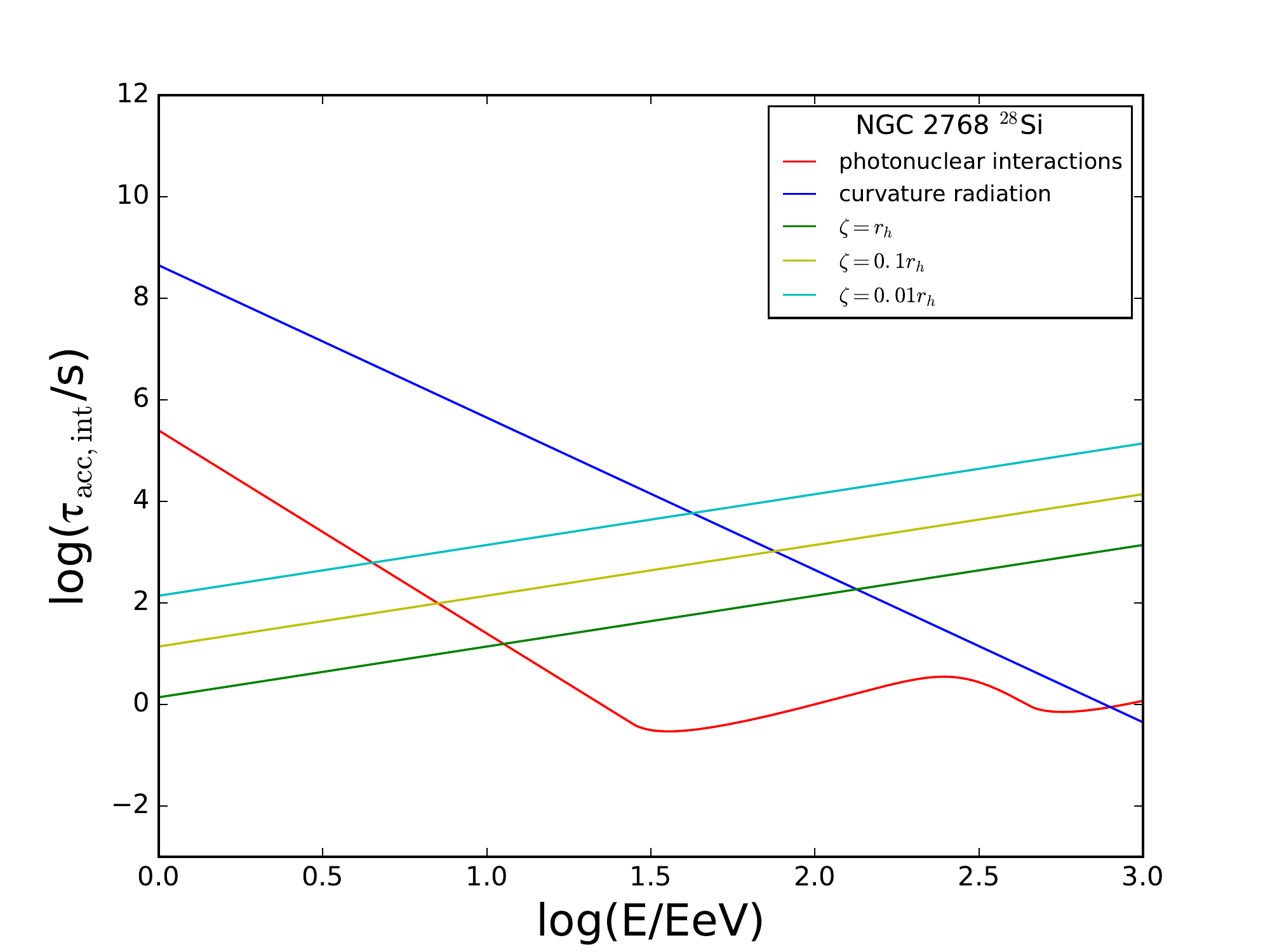}{0.99} 
\end{minipage} 
\caption{Comparison of acceleration and interaction time scales for
  NGC 2768. The
  bumpy curve indicates the interaction time scale for photonuclear
  interactions. The straight line with  negative slope indicate the
  time scale due to curvature radiation. The straight lines with
  positive slope correspond to acceleration time scales with different
$\zeta$. The upper panels correspond to $\beta = -4$ and the lower
panels to $\beta = -5$.}
\label{fig:ngc2768}
\end{figure*}

\begin{figure*}[tpb] 
\begin{minipage}[t]{0.49\textwidth}
    \postscript{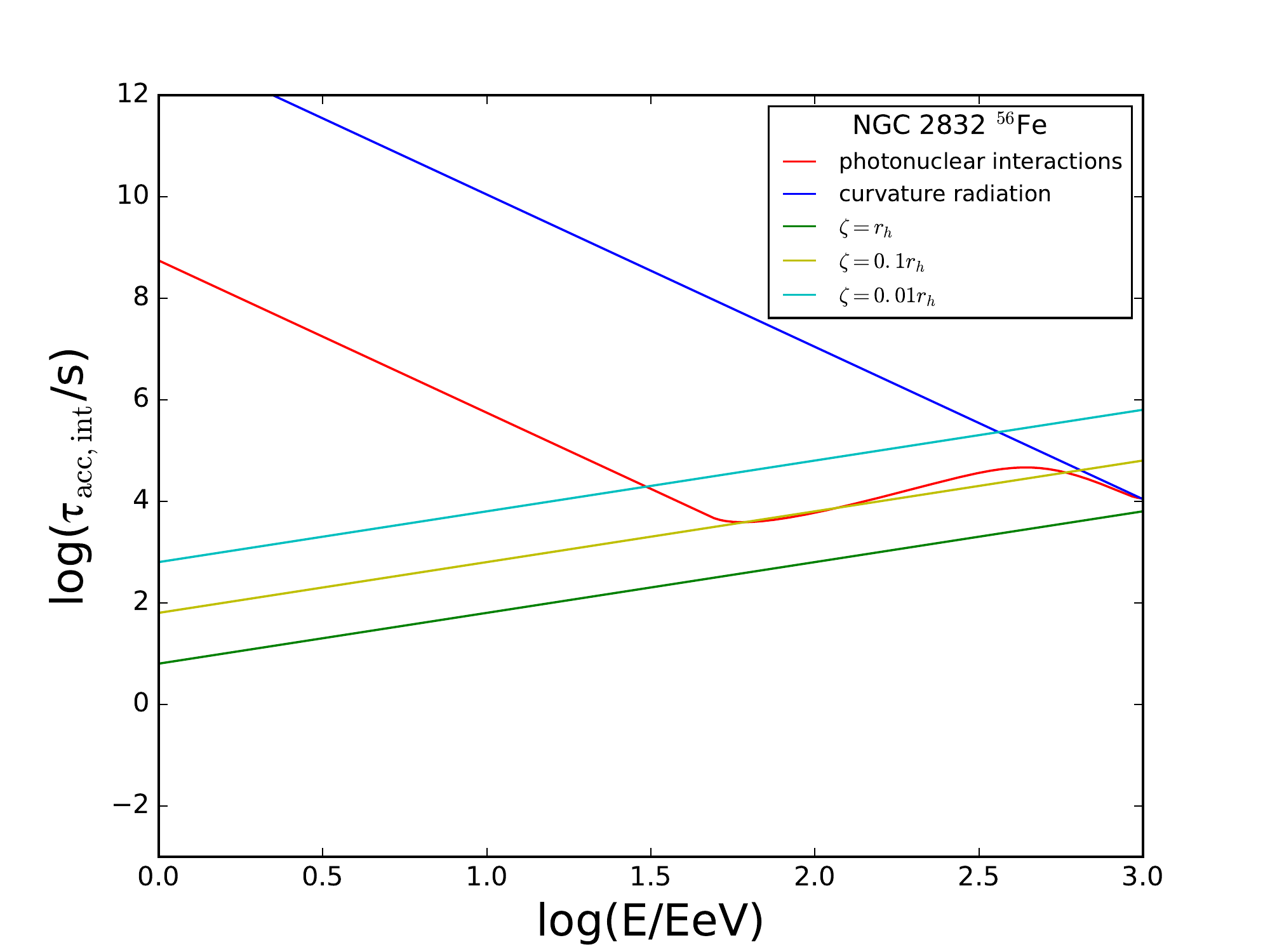}{0.99} 
\end{minipage} 
\begin{minipage}[t]{0.49\textwidth}
    \postscript{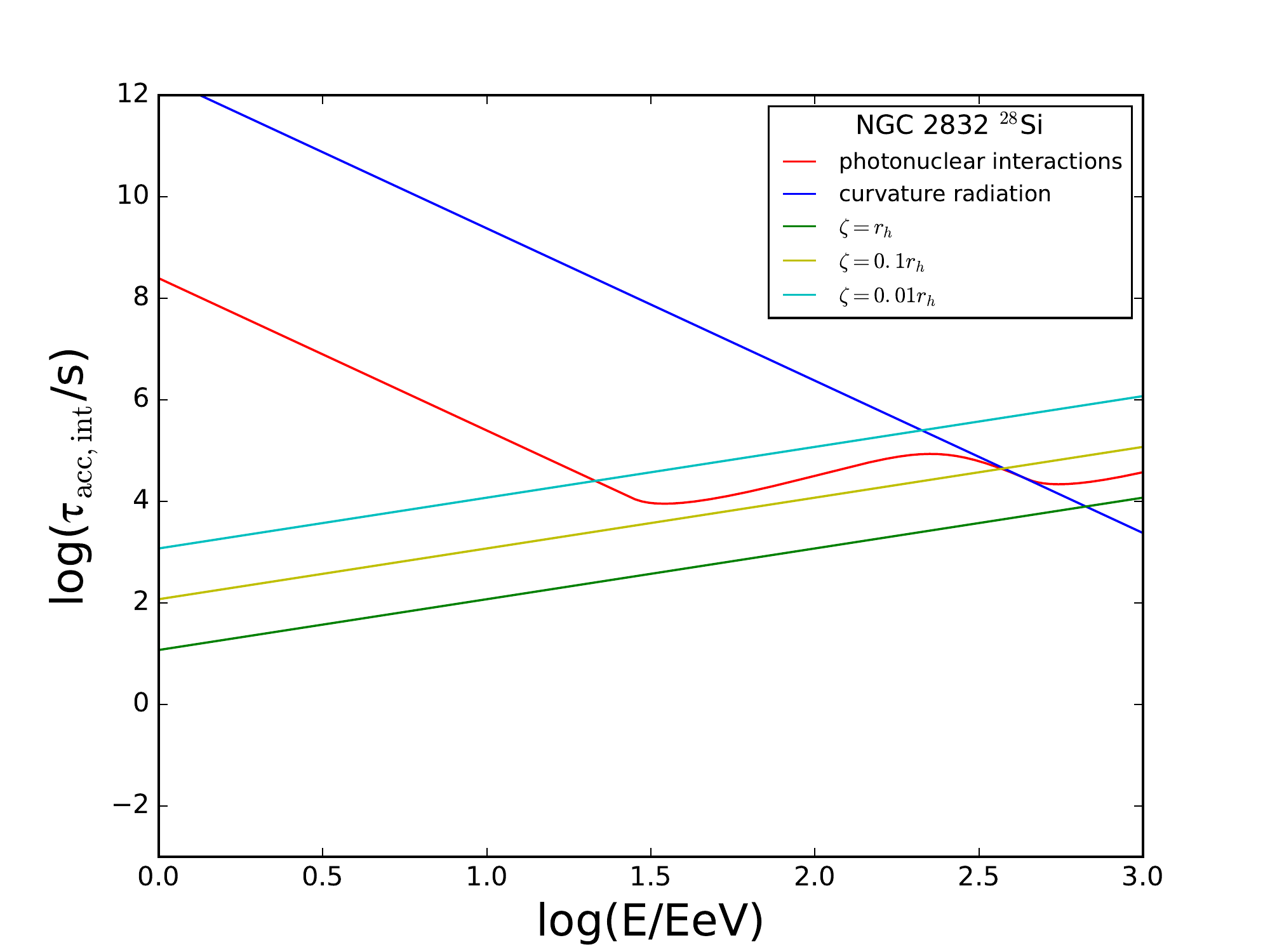}{0.99} 
\end{minipage}
\hfill \begin{minipage}[t]{0.49\textwidth}
  \postscript{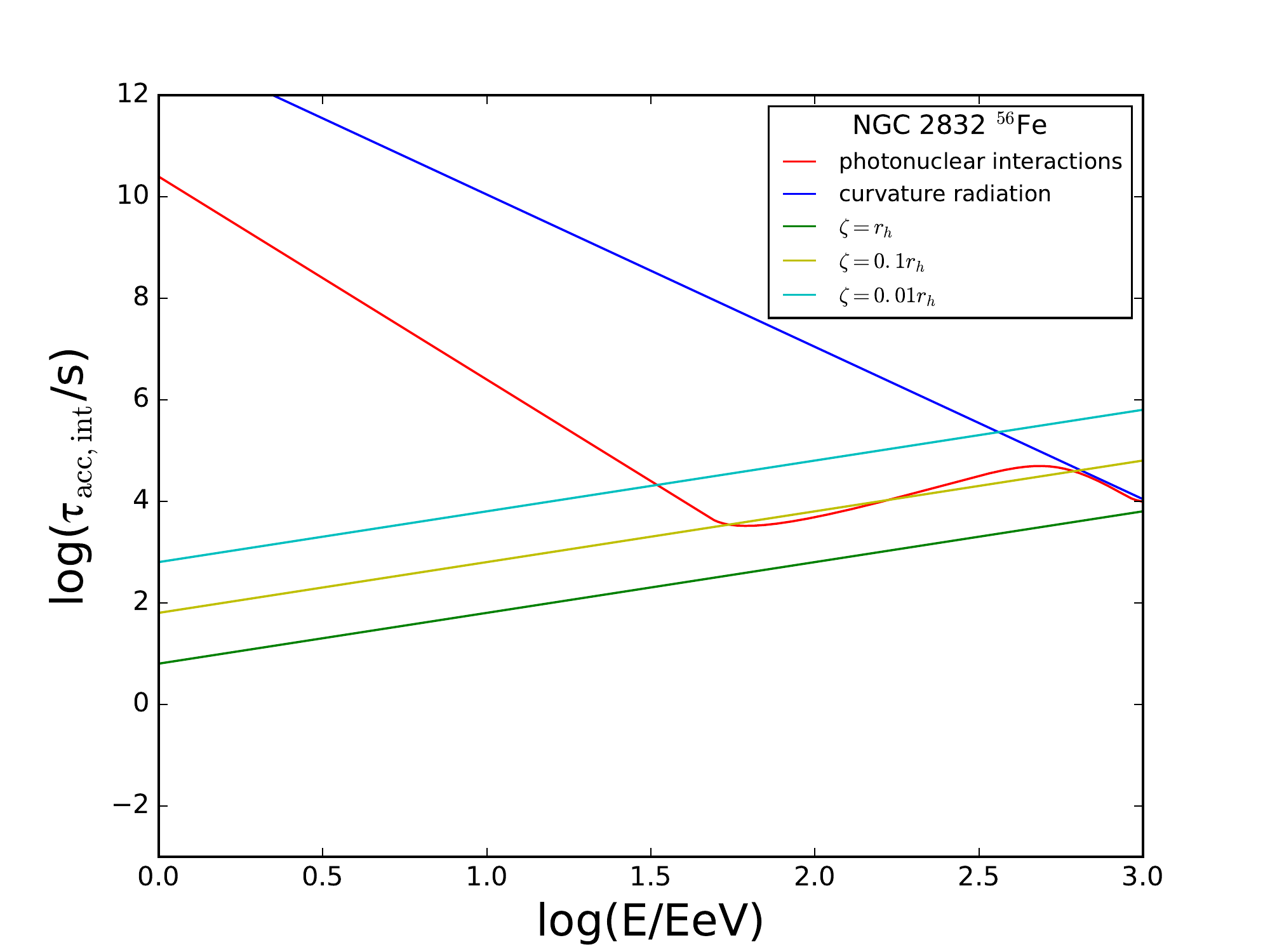}{0.99} 
\end{minipage} 
\begin{minipage}[t]{0.49\textwidth}
  \postscript{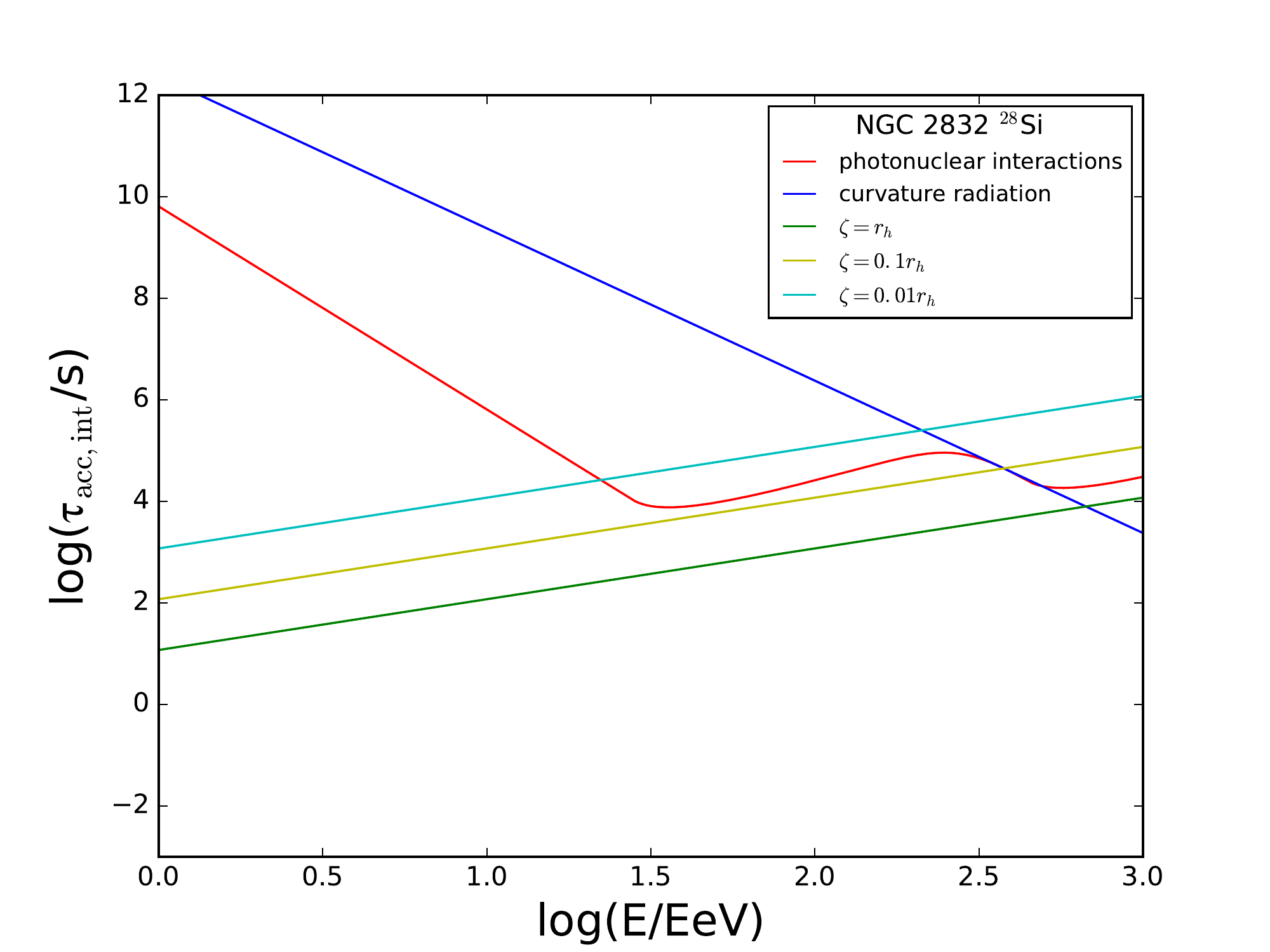}{0.99} 
\end{minipage} 
\caption{Comparison of acceleration and interaction time scales for
  NGC 2832. Conventions are as in Fig.~\ref{fig:ngc2768}.}
\label{fig:ngc2832}
\end{figure*}

\begin{figure*}[tbp] 
\begin{minipage}[t]{0.49\textwidth}
    \postscript{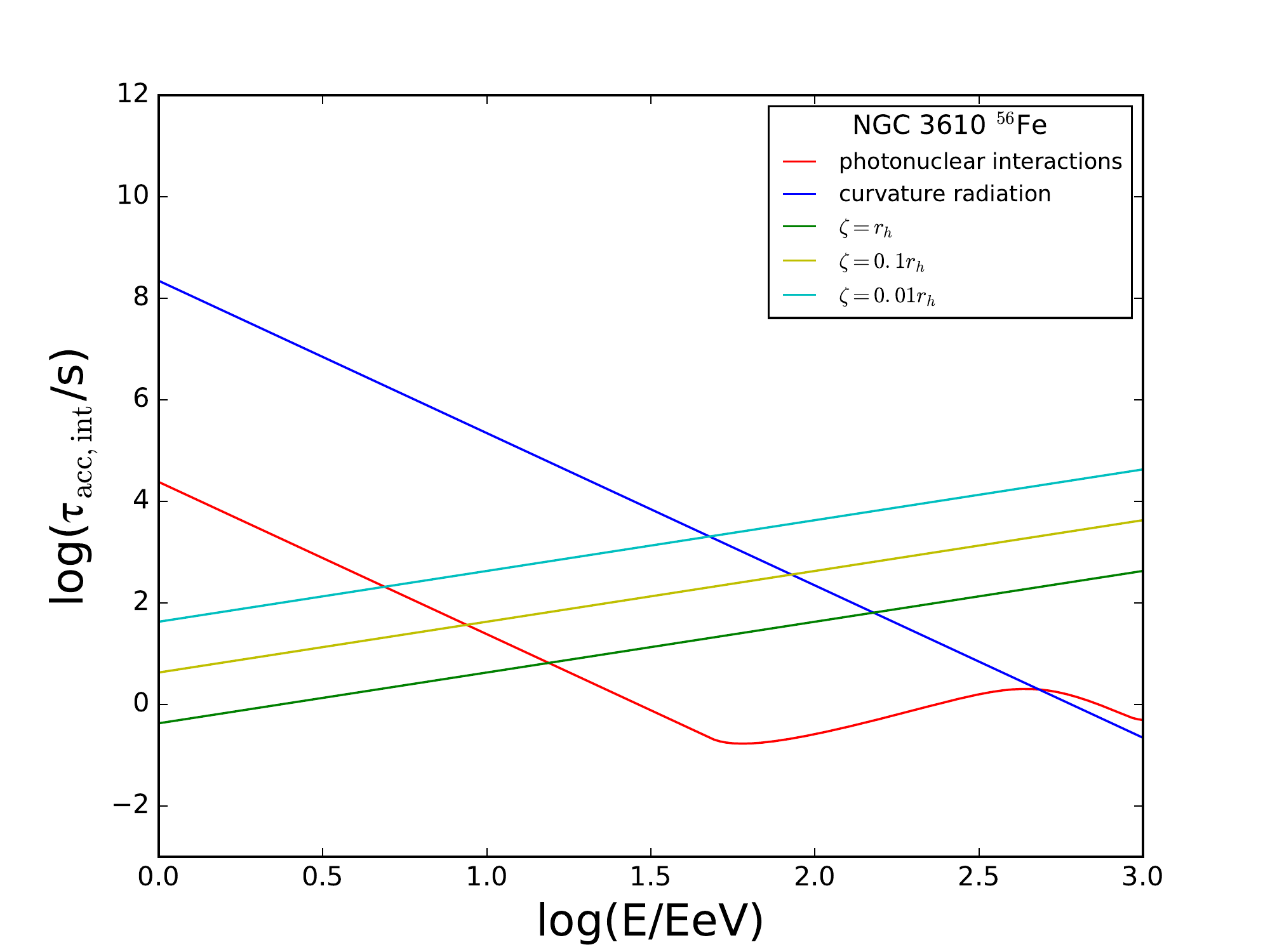}{0.99} 
\end{minipage} 
\begin{minipage}[t]{0.49\textwidth}
    \postscript{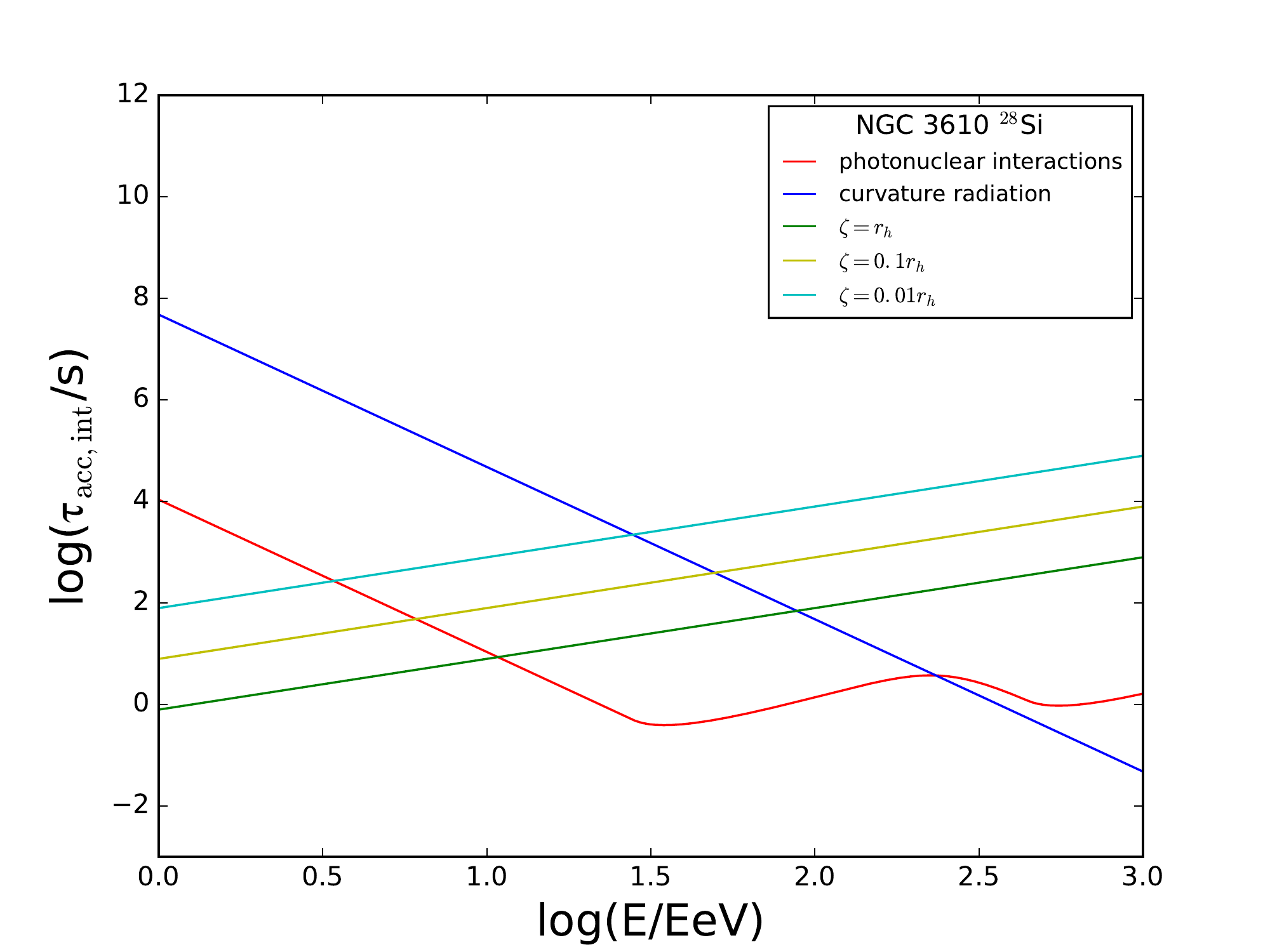}{0.99} 
\end{minipage}
\hfill \begin{minipage}[t]{0.49\textwidth}
  \postscript{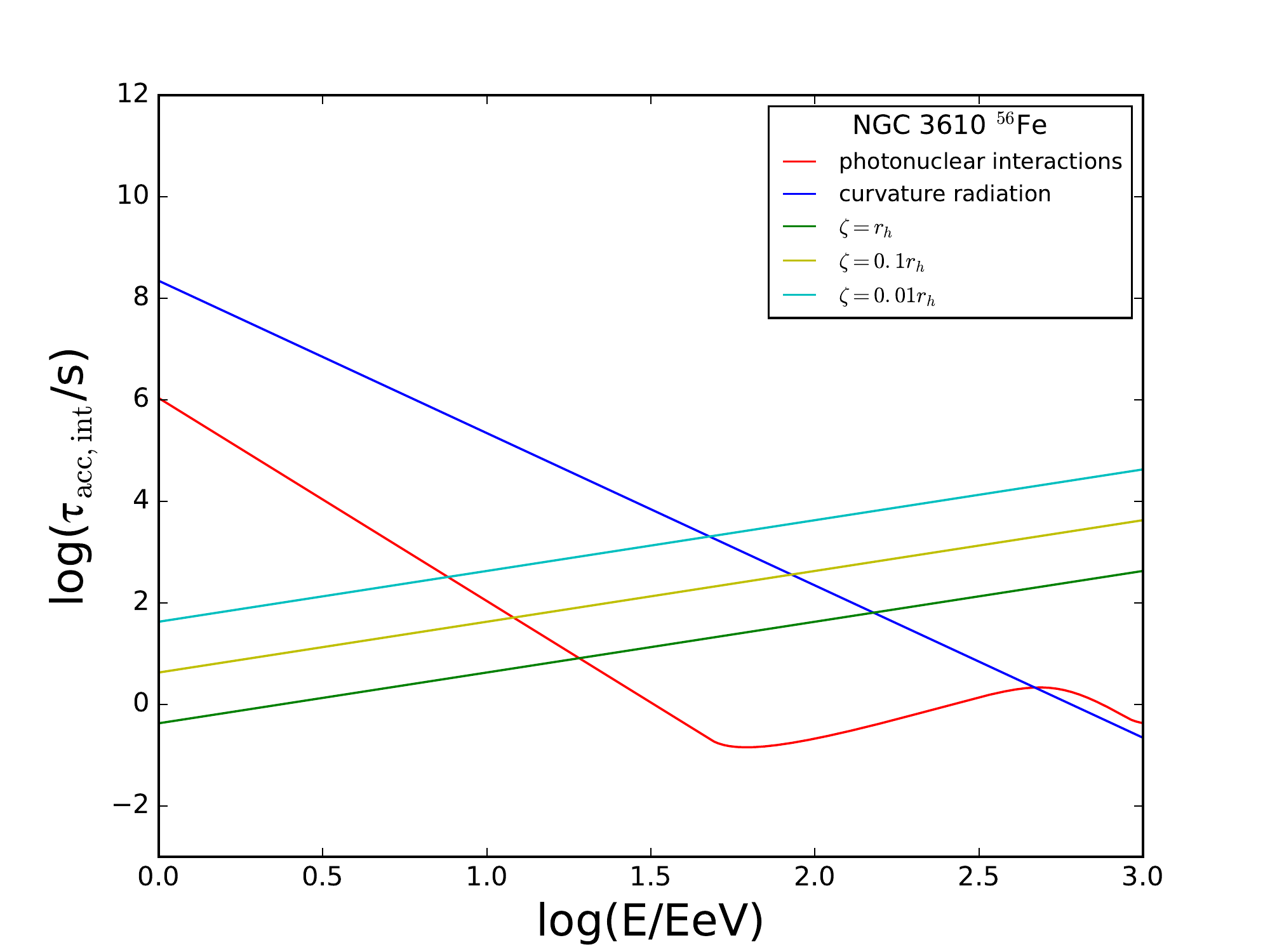}{0.99} 
\end{minipage} 
\begin{minipage}[t]{0.49\textwidth}
  \postscript{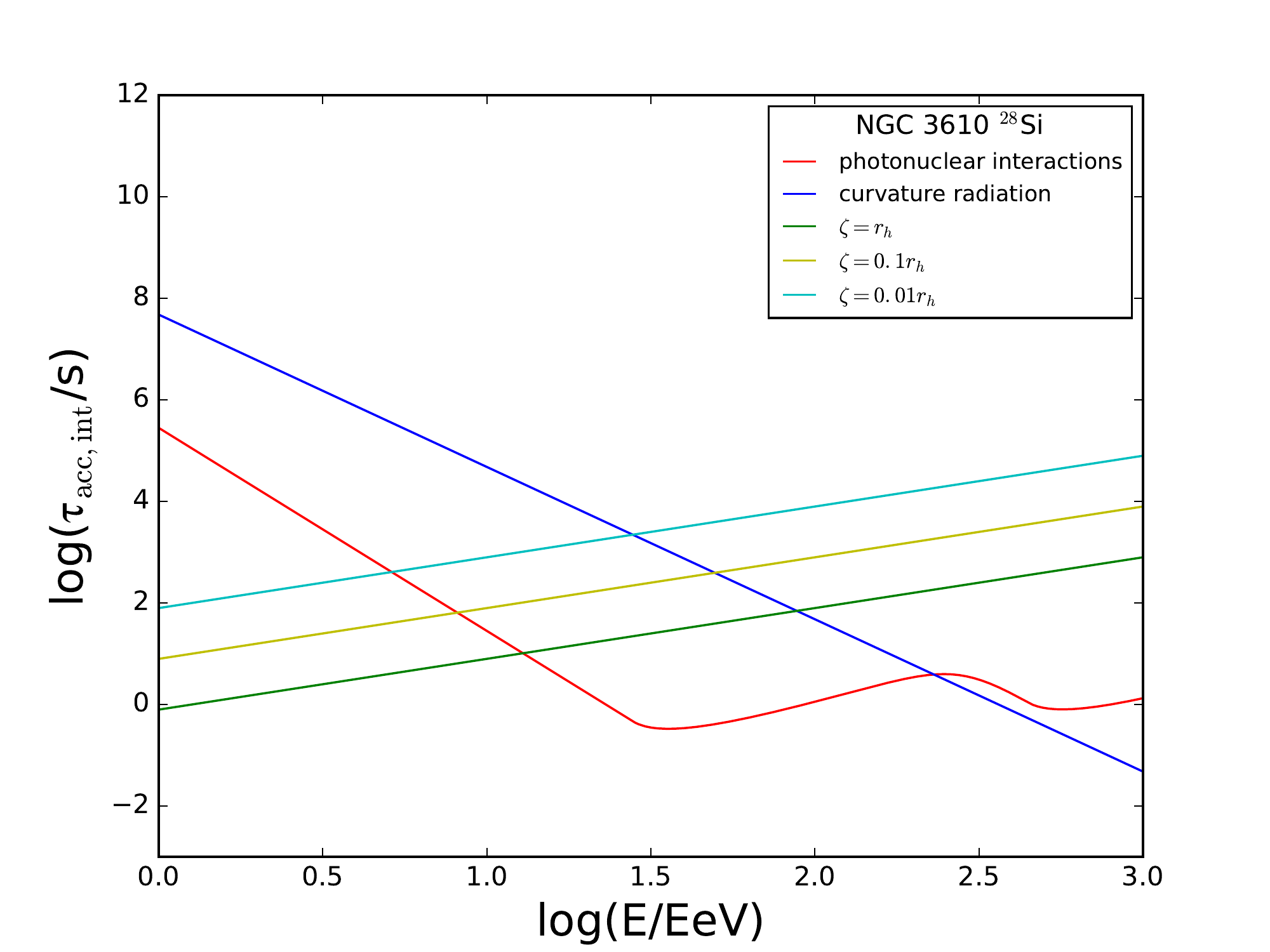}{0.99} 
\end{minipage} 
\caption{Comparison of acceleration and interaction time scales for
  NGC 3610. Conventions are as in Fig.~\ref{fig:ngc2768}.}
\label{fig:ngc3610}
\end{figure*}

\begin{figure*}[tbp] 
\begin{minipage}[t]{0.49\textwidth}
    \postscript{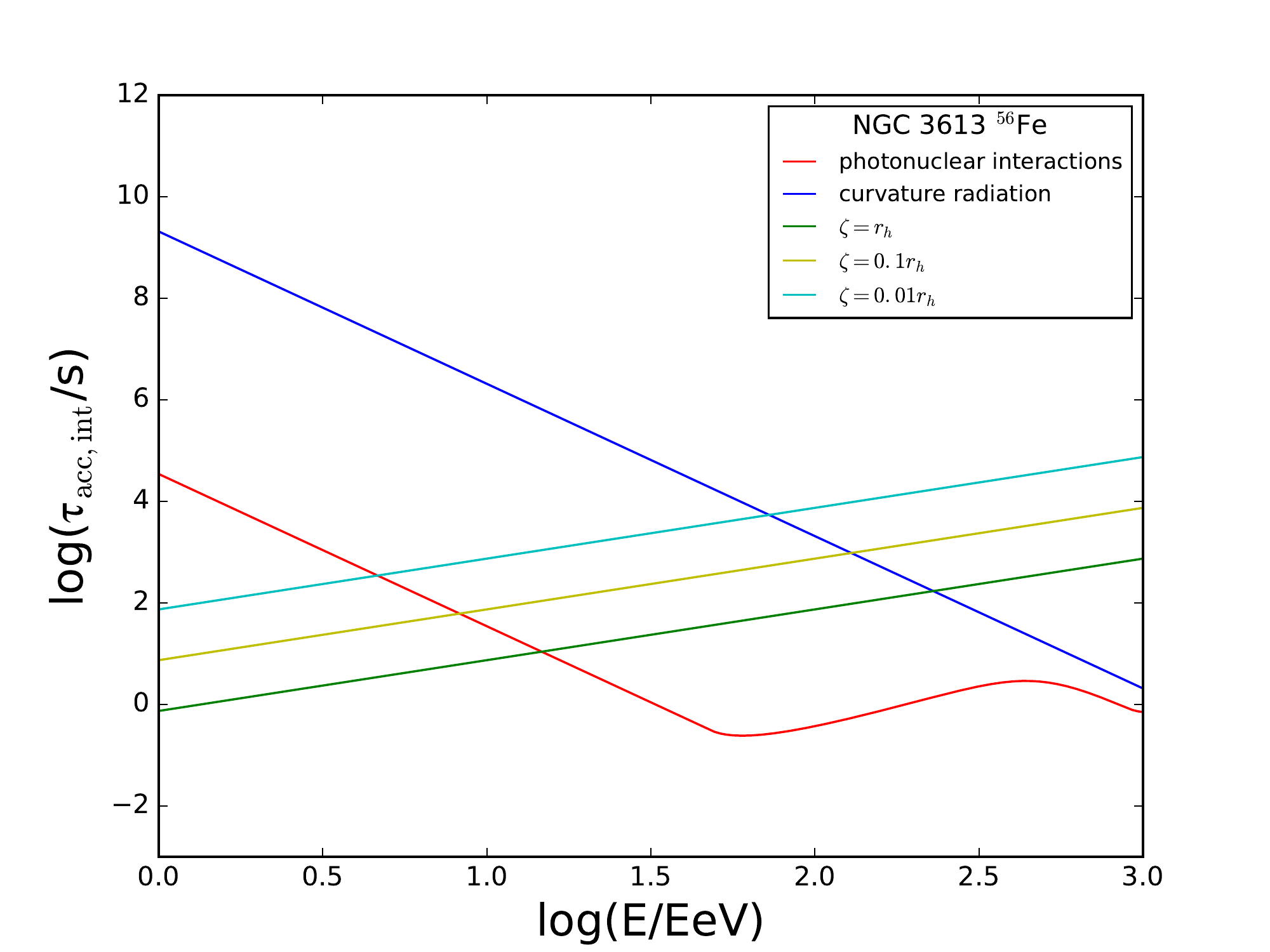}{0.99} 
\end{minipage} 
\begin{minipage}[t]{0.49\textwidth}
    \postscript{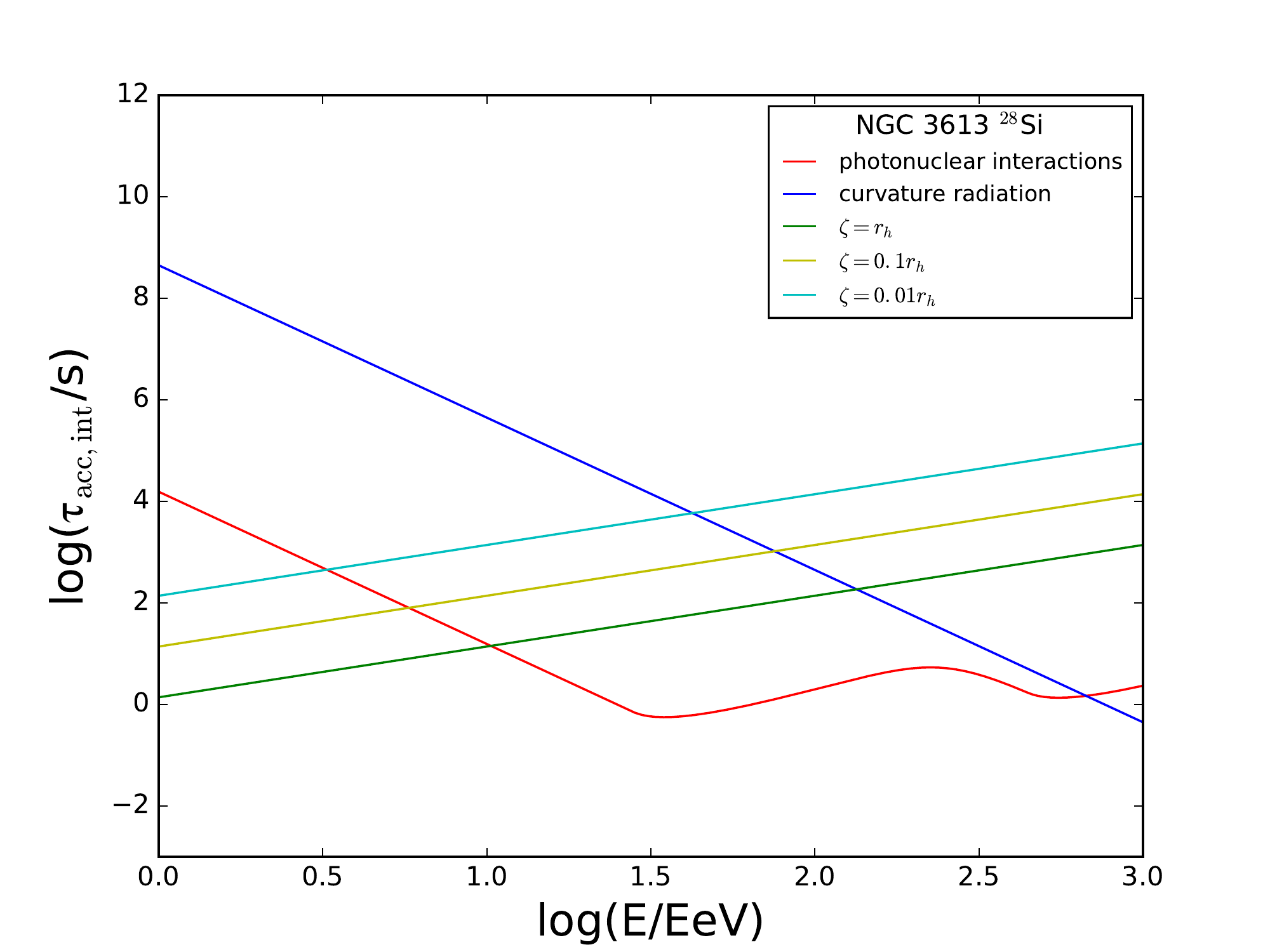}{0.99} 
\end{minipage}
\hfill \begin{minipage}[t]{0.49\textwidth}
  \postscript{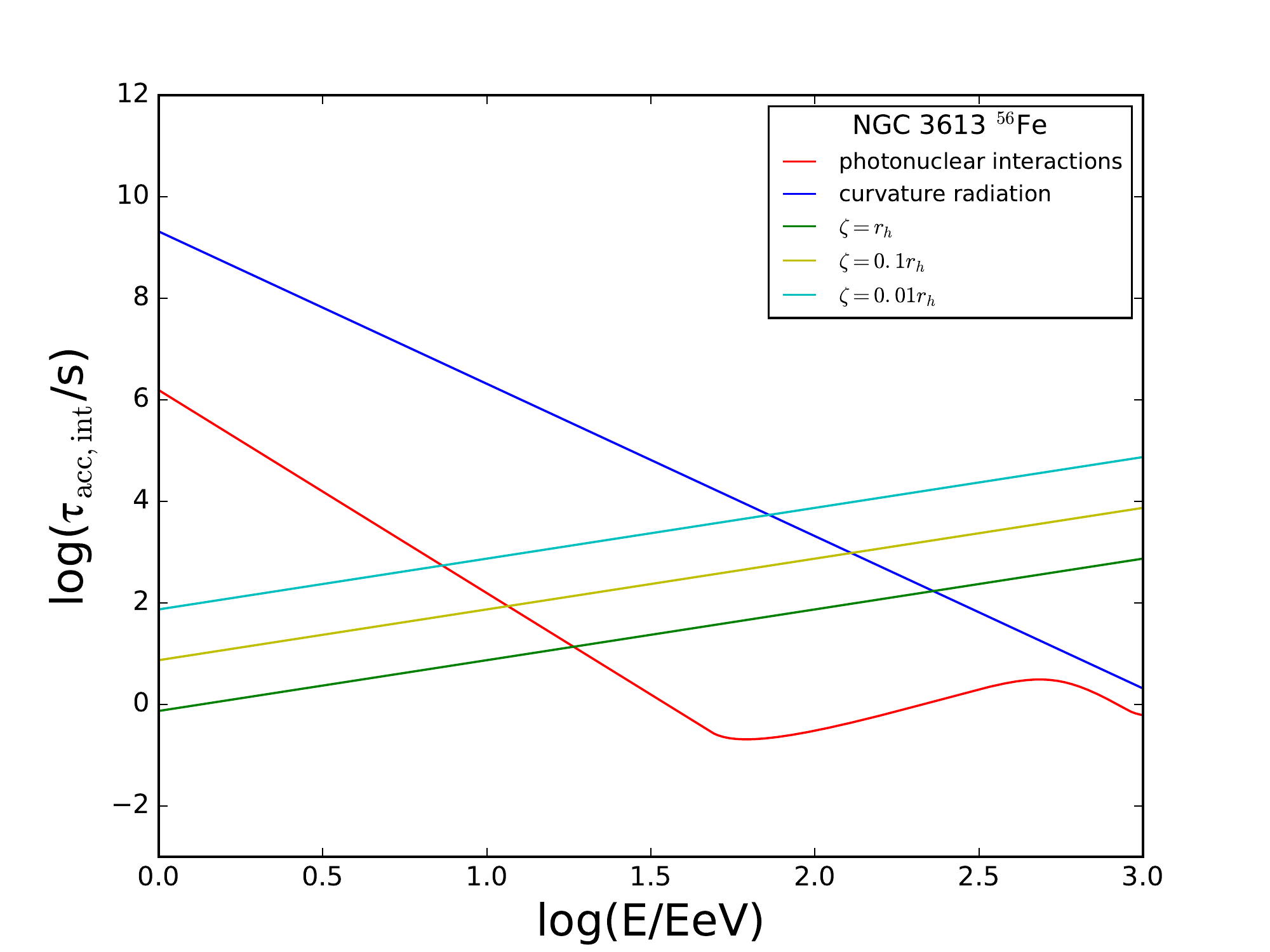}{0.99} 
\end{minipage} 
\begin{minipage}[t]{0.49\textwidth}
  \postscript{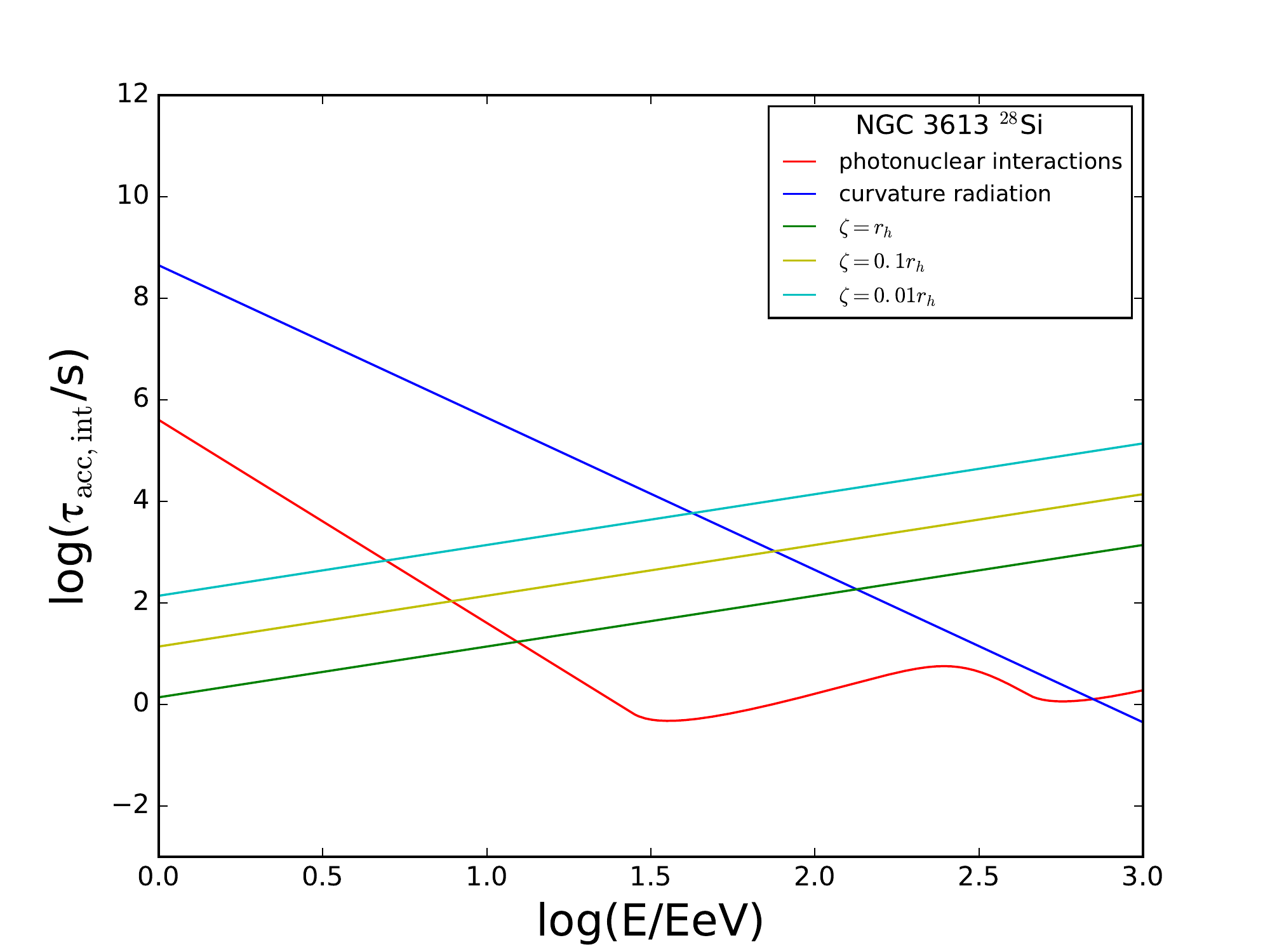}{0.99} 
\end{minipage} 
\caption{Comparison of acceleration and interaction time scales for
  NGC 3613. Conventions are as in Fig.~\ref{fig:ngc2768}.}
\label{fig:ngc3613}
\end{figure*}

\begin{figure*}[tbp] 
\begin{minipage}[t]{0.49\textwidth}
    \postscript{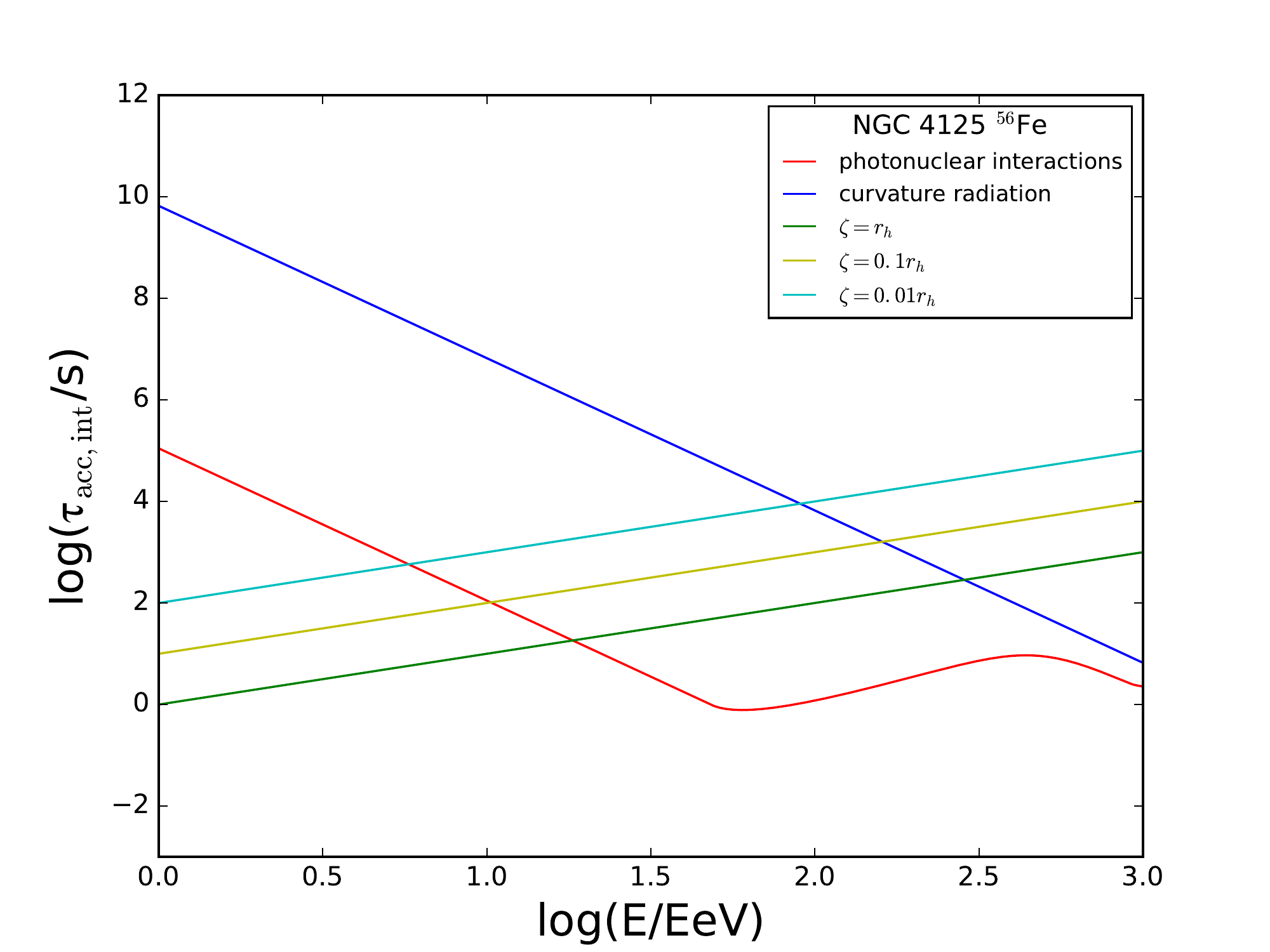}{0.99} 
\end{minipage} 
\begin{minipage}[t]{0.49\textwidth}
    \postscript{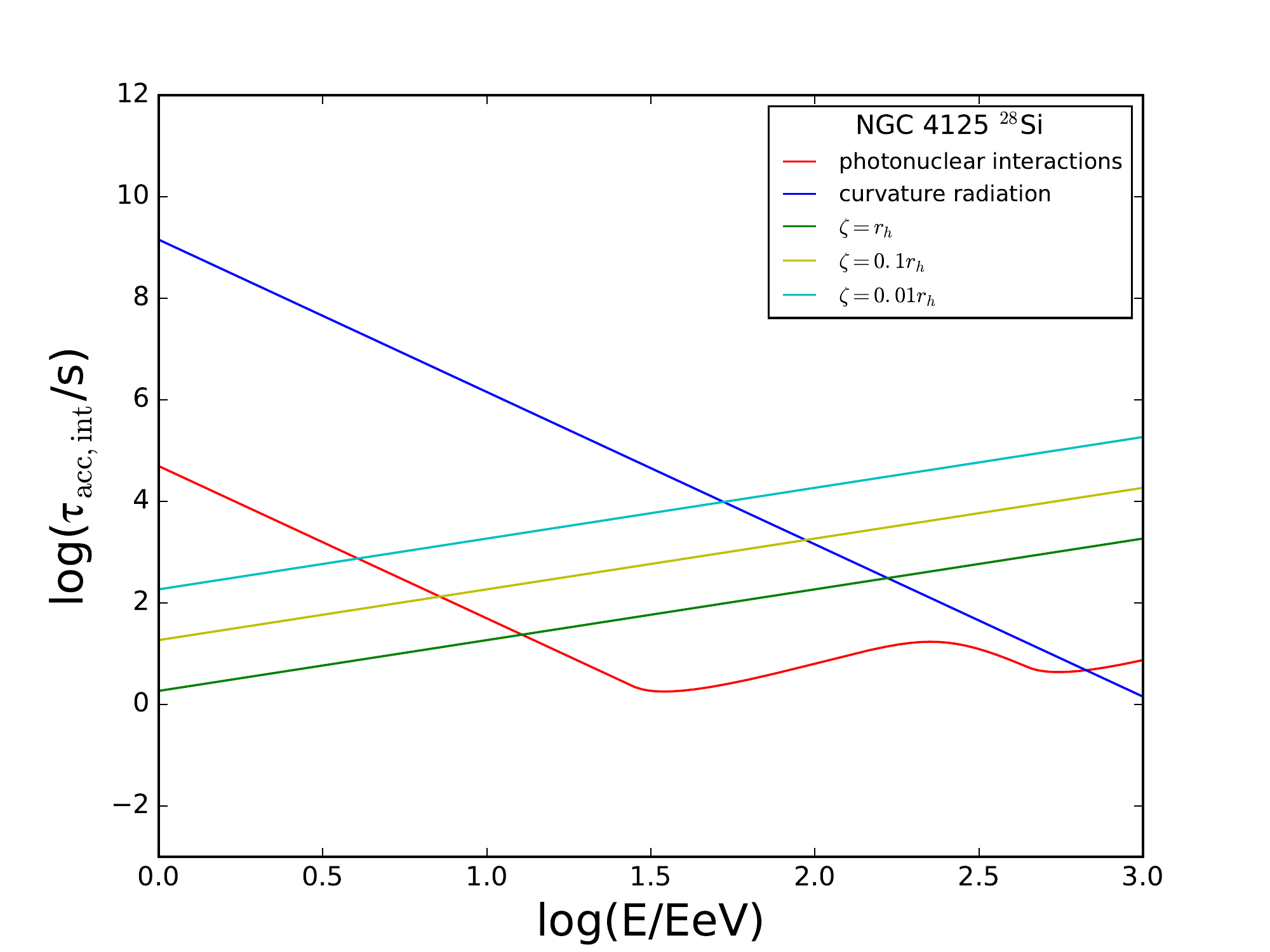}{0.99} 
\end{minipage}
\hfill \begin{minipage}[t]{0.49\textwidth}
  \postscript{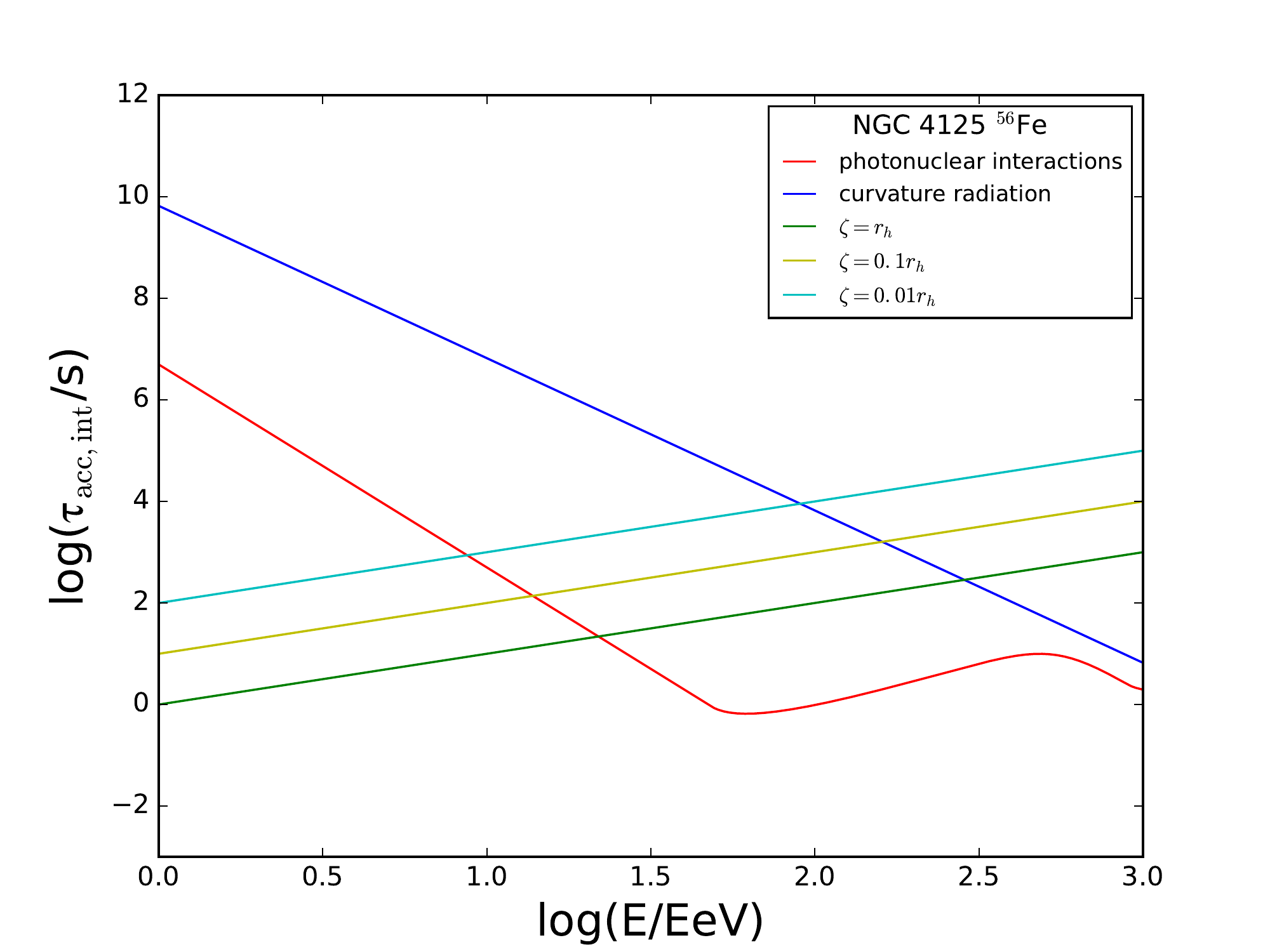}{0.99} 
\end{minipage} 
\begin{minipage}[t]{0.49\textwidth}
  \postscript{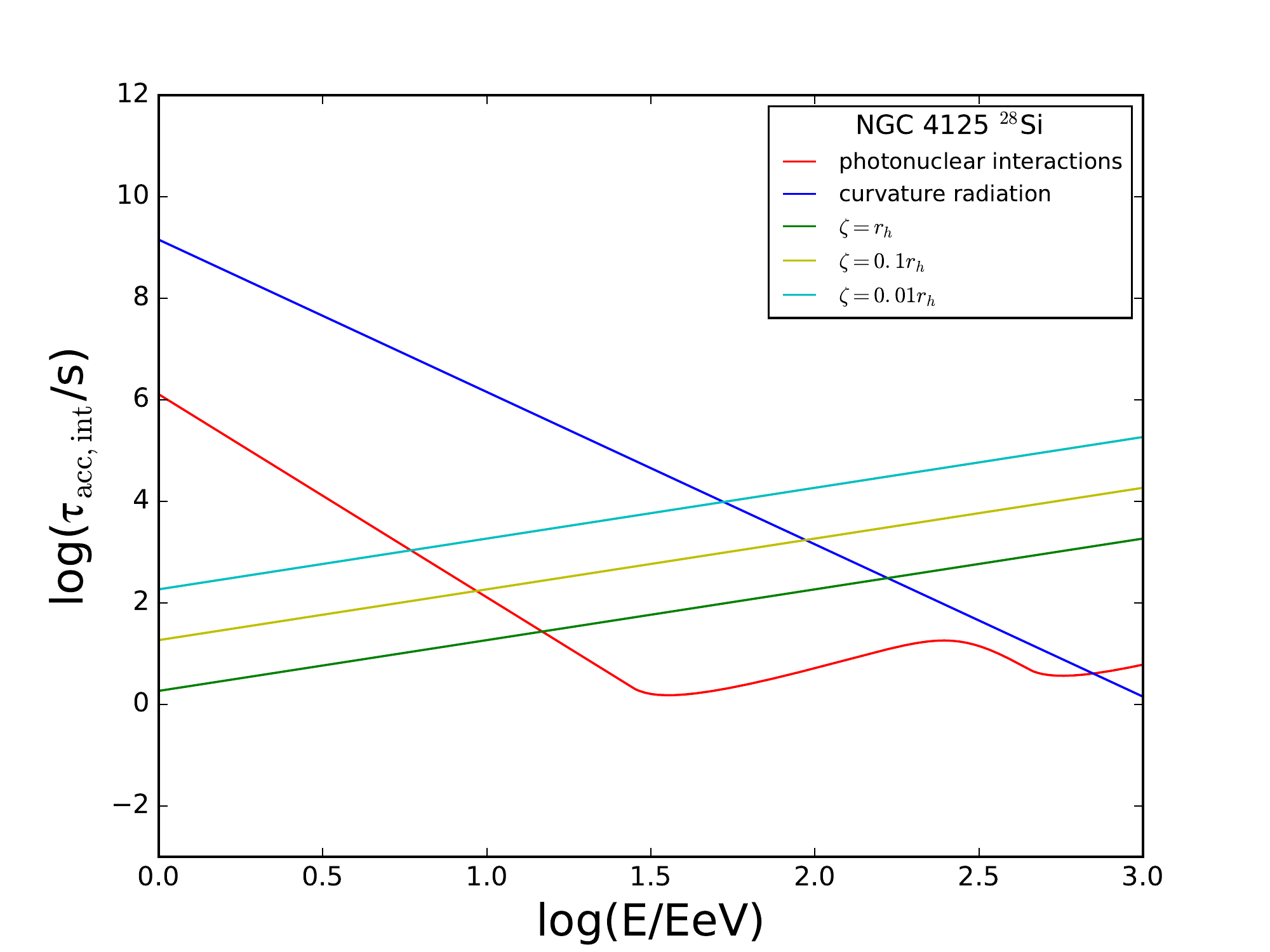}{0.99} 
\end{minipage} 
\caption{Comparison of acceleration and interaction time scales for
  NGC 4125. Conventions are as in Fig.~\ref{fig:ngc2768}.}
\label{fig:ngc4125}
\end{figure*}

\begin{figure*}[tpb] 
\begin{minipage}[t]{0.49\textwidth}
    \postscript{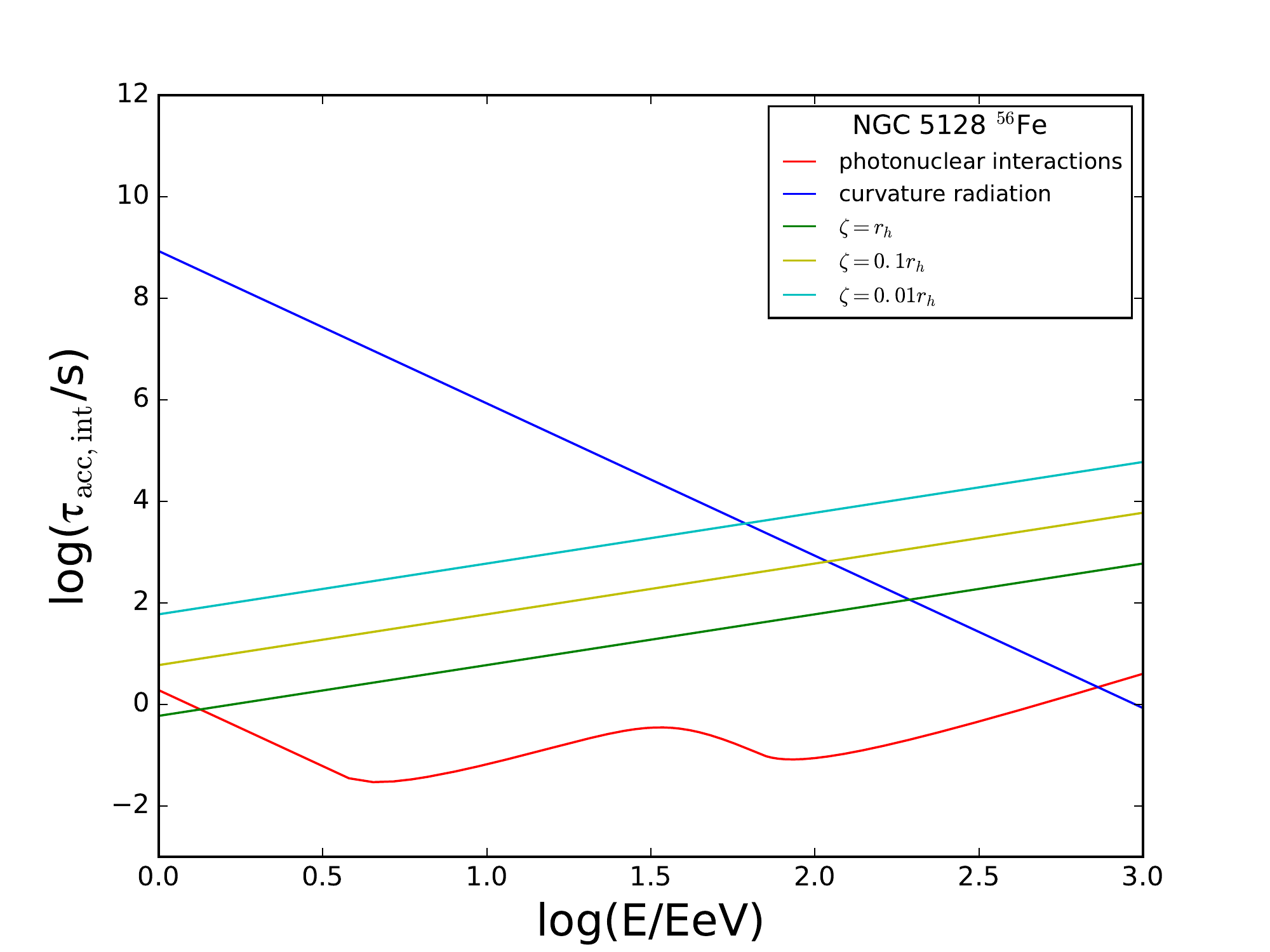}{0.99} 
\end{minipage} 
\begin{minipage}[t]{0.49\textwidth}
    \postscript{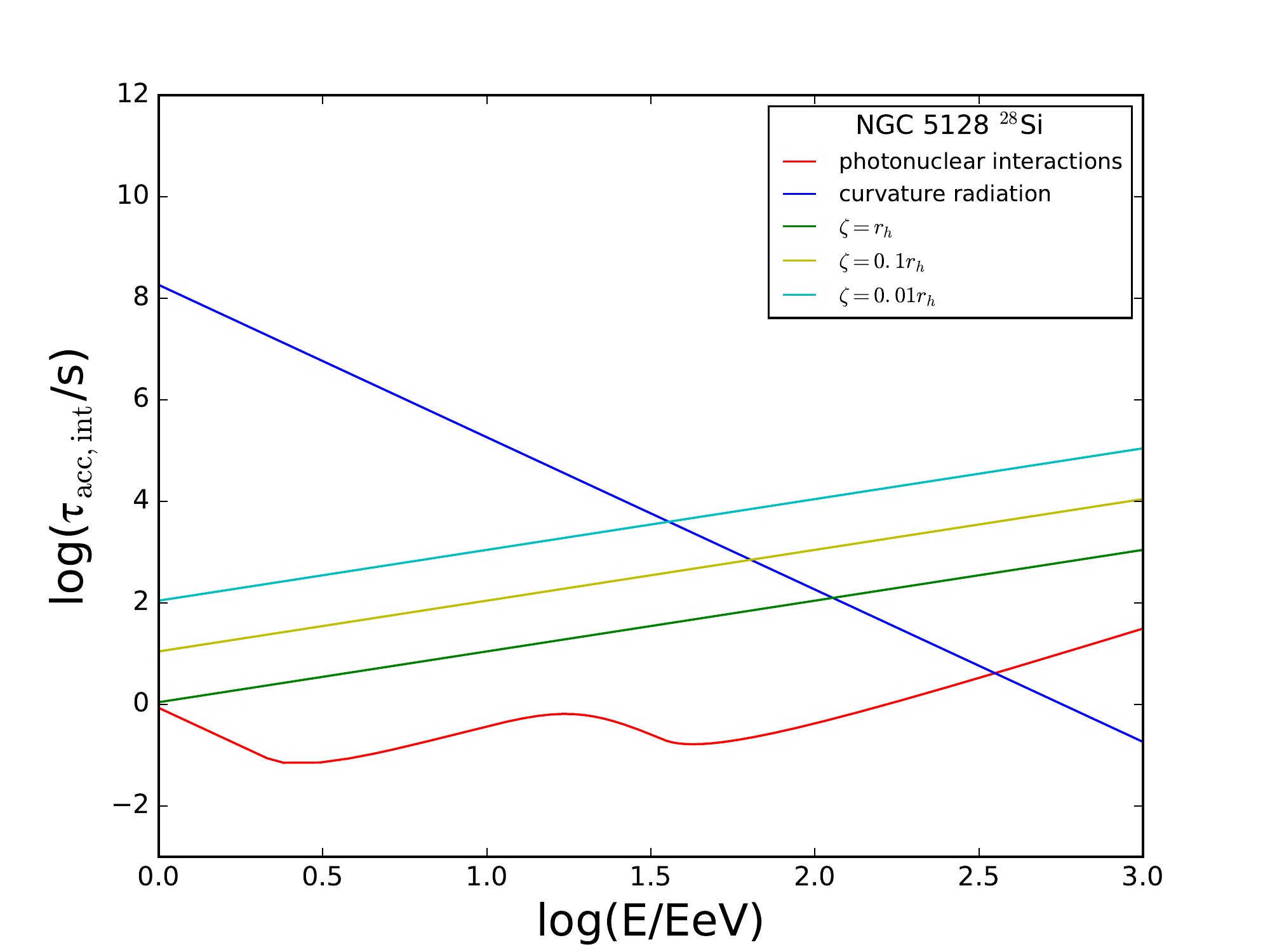}{0.99} 
\end{minipage}
\hfill \begin{minipage}[t]{0.49\textwidth}
  \postscript{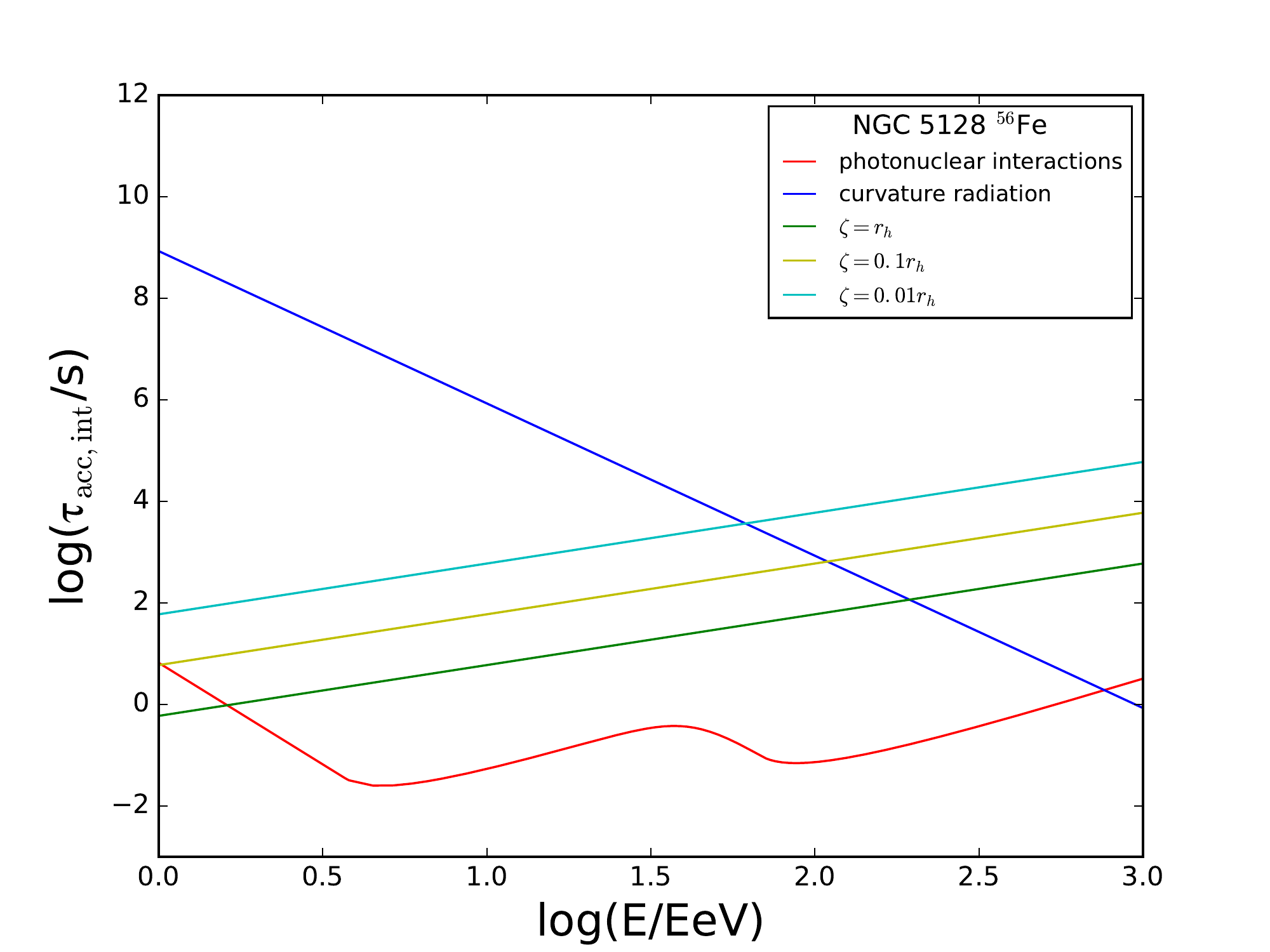}{0.99} 
\end{minipage} 
\begin{minipage}[t]{0.49\textwidth}
  \postscript{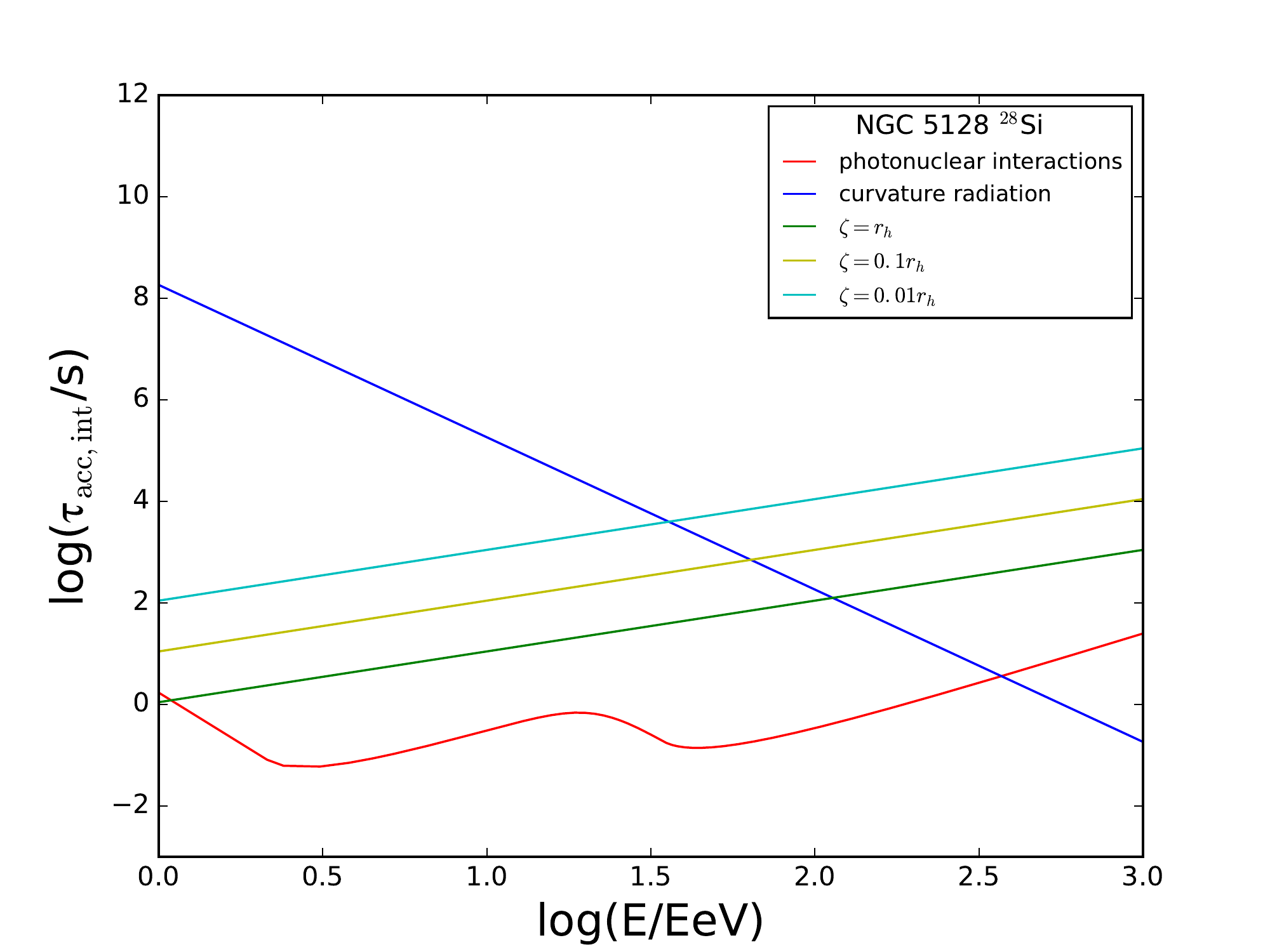}{0.99} 
\end{minipage} 
\caption{Comparison of acceleration and interaction time scales for
  NGC 5128. Conventions are as in Fig.~\ref{fig:ngc2768}.}
\label{fig:ngc5128}
\end{figure*}

In the absence of energy losses the maximum  Lorentz factor is given by
\begin{equation}
\gamma_{\rm max}^{\rm acc} \sim 10^{11} \frac{Z}{A} \ M_9 \, B_4 \
\left(\frac{\zeta}{r_h} \right)^2 ,
\label{gamma-cr}
\end{equation}
where $E = \gamma A m_p$, with $m_p$ the proton mass and  $A$ the nuclear baryon number.
Within the potential drop, the nuclei  follow the curved magnetic
field lines and so emit curvature-radiation photons. The energy loss
rate or total power radiated away by a single cosmic ray is~\cite{Ochelkov} 
\begin{equation}
\left. \frac{d E}{dt} \right|_{\rm loss}^{\rm rad} = \frac{2}{3} \frac{Z^2 e^2 c}{r_c^2}
\gamma^4 \,,
\end{equation}
where $r_c$ is the curvature radius of the magnetic field lines, and
$\gamma$ is the Lorentz factor of the radiating particles. 
The characteristic cooling timescale is then given by
\begin{equation}
\tau_{\rm int}^{\rm rad}  =  \gamma A m_p  \left[\left. \frac{d
      E}{dt} \right|_{\rm loss}^{\rm rad} \right]^{-1}   \, .
\end{equation}
Acceleration gains are balanced by radiative losses. In the
absence of other damping mechanisms,  the radiation reaction limit
is~\cite{Levinson:2000nx,Levinson:2002ea} 
\begin{equation}
\gamma_{\rm max}^{\rm rad} \sim 10^{11} \left(\frac{B_4 }{Z} \right)^{1/4} M_9^{1/2}
\left(\frac{r_c}{r_h} \right)^{1/2}
\left(\frac{\zeta}{r_h}\right)^{1/4} \, .
\end{equation}
All in all, direct electric field acceleration on the black hole magnetosphere would allow Lorentz factors up to
\begin{equation}
\gamma_{\rm max} = {\rm min} \, \left\{\gamma_{\rm max}^{\rm acc}, \
\gamma_{\rm max}^{\rm rad} \right\}\, .
\end{equation}
Note that for our fiducial values $B_4 \sim 1$, $M_9 \sim 1$, and $r_c
\sim r_h$, we have $\gamma_{\rm max}^{\rm acc} \sim \gamma_{\rm
  max}^{\rm rad}$ if $\zeta \sim r_h$. However, by taking $\zeta <
r_h$ one can always suppress the emission of curvature photons.
Moreover, as we show in what follows photonuclear interactions on the
ambient photon background become the dominant mechanism for energy losses.

 The estimation of the photon density in the vicinity of
  the SMBH is not straightforward. Studies for Cen A are quite
  extensive~\cite{Whysong:2002ks,Meisenheimer:2007zv}, and the photon
  spectrum around the SMBH is consistent with a thermal origin: mostly dust emission
  heated by the nucleus~\cite{Whysong:2002ks}. For simplicity, in our
  calculations we approximate the photon spectrum as a broken
  power-law,
\begin{equation}
      n(\varepsilon) = n_0
        \begin{cases}
           (\varepsilon/\varepsilon_0)^{\, \alpha} & \quad \varepsilon < \varepsilon_0 \\
           (\varepsilon/\varepsilon_0)^{\, \beta} & \quad \text{otherwise} \, ,
        \end{cases}
\label{app:eq:photonfield}
\end{equation}
which is also consistent with the data. Here $\varepsilon$ is the
photon energy and the maximum of the density is at an energy of
$\varepsilon_0$. The broken power-law spectrum allows a complete
analytic treatment of photonuclear interactions, and the global result
does not depend on the exact shape of the photon spectrum.  Namely,
the interaction times are comparable if the photon density is assumed
to follow a (modified) black body spectrum~\cite{Unger:2015laa}.  For more distance
sources, it is not trivial to disentangle which photons are near the
SMBH. As a first approximation, we use the galaxy-wide average far
infra-red (FIR) flux reported in the NASA extragalactic database to
normalize the spectrum and estimate the spectral indices~\cite{NED}. If a
correlation becomes evident in the future a more detailed analysis
would be required. By averaging over the thermal spectra of the
sources studied in~\cite{Wiklind} we take $\alpha = 2$ and
\mbox{$-5 \leq \beta \leq -4$.} 

 We normalize the spectrum to the
FIR luminosity in the vicinity of the SMBH
\begin{eqnarray}
L_{\rm FIR}^{\rm BH} & = & 4\pi r_g^2c\int
n(\varepsilon)\,\varepsilon\,d\varepsilon \nonumber \\
& = & 4\pi\,r_g^2\,c\,n_0\left\{\int_{\varepsilon_{\rm
      min}}^{\varepsilon_0} \! \! \!
  (\varepsilon/\varepsilon_0)^\alpha\,\varepsilon\,d\varepsilon+\int_{\varepsilon_0}^\infty
  \! \! \! (\varepsilon/\varepsilon_0)^\beta\,\varepsilon\,d\varepsilon\right\} \!,
\end{eqnarray}
where $\varepsilon_{\rm min} = 10^{-4}~{\rm eV}$ and $\varepsilon_0 =
10^{-2}~{\rm eV}$. There is a general consensus that the extended FIR
emission along the dust lane of Cen A, $L_{\rm FIR} \sim 2 \times
10^{10} L_\odot$, cannot only be attributed to dust heated by the
active nucleus~\cite{Marshall:1988,Eckart,Wild:2000hq,Stawarz:2006ku}. Optical detections of luminous
HII regions in relatively unoscured regions of the dark
lane~\cite{Graham:1981} suggest that recently formed stars embedded in
the dust lane must be responsible for most of the broadly extended FIR
emission. Only about 1\% of the FIR emission comes from the active
nucleus. With this in mind, we take $L_{\rm FIR}^{\rm BH} /L_{\rm FIR}
\sim 10^{-2}$. The normalization factor is then
\begin{equation}
n_0=\frac{L_{\rm FIR}^{\rm BH}}{4\pi\,r_g^2\,c} 
  \left\{\frac{\varepsilon_0^{-\alpha}}{(\alpha+2)}\left[\varepsilon_0^{\alpha+2}-\varepsilon_{\rm
        min}^{\alpha+2}\right] -
    \frac{\varepsilon_0^2}{(\beta+2)} \right\}^{-1} \, .
\end{equation}

In Table~\ref{tabla} we summarize the relevant
properties of a subgroup of 5 candidate sources shown in
Fig.~\ref{fig:1}. In the case of no-detection in the FIR we normalize the
photon flux to the reported upper-limit. To determine the distance to
these sources we adopt the usual concordance cosmology of a flat
universe dominated by a cosmological constant, with $\Omega_\Lambda =
0.692 \pm 0.012$ and a cold dark matter plus baryon component
$\Omega_m = 0.308 \pm 0.012$; the Hubble parameter as a function of
redshift is given by $H^2(z) = H_0^2[\Omega_m (1 + z)^3 +
\Omega_\Lambda]$, normalized to its value today, $H_0 = 100~h \, {\rm
  km} \, {\rm s}^{-1} \, {\rm Mpc}^{-1}$, with $h =
0.678$~\cite{Olive:2016xmw}.

To characterize the population of nearby quasar remnants we also study
the case of NGC 5128. This case is special, as the photon spectrum
peaks in the mid-infrared. We take $\epsilon_0 = 0.13~{\rm eV}$ and
$L_{\rm mid-IR}/L_\odot = 1.3 \times 10^8$. The mass of the SMBH is
somewhat uncertain~\cite{Cappellari:2008db,Neumayer:2007ic,Marconi:2005qx}. In
our calculations we adopt $M_9 = 0.1$, yielding $n_0^{-4}=3.0 \times
10^{23}~{\rm MeV^{-1} \, cm^{-3}}$ and $n_0^{-5}=3.9 \times
10^{23}~{\rm MeV^{-1} \, cm^{-3}}$.

The interaction time for a highly relativistic nucleus propagating
through an isotropic photon background with energy $\varepsilon$ and spectrum
$n(\varepsilon)$, normalized so that the total number of photons in a
box is $\int n ( \varepsilon ) d \varepsilon$, is given by~\cite{Stecker:1969fw}
\begin{equation}
 \frac{1}{\tau_\mathrm{int}^{A\gamma}} = \frac{c}{2} \,\int_0^\infty
                 d\varepsilon \,\frac{n(\varepsilon)}{\gamma^2 \varepsilon^2}\, \int_0^{2\gamma\varepsilon}
                 d\varepsilon^\prime \, \varepsilon^\prime\, \sigma(\varepsilon^\prime),
\label{app:eq:interaction}
\end{equation}
where $\sigma(\varepsilon^\prime)$ is the photonuclear interaction
cross section of a nucleus by a photon energy
$\varepsilon'$ in the rest frame of the nucleus.

We have found that for the considerations in the present work, the cross section can be safely
approximated by the single pole of the narrow-width approximation,
\begin{equation}
\sigma (\varepsilon') = \pi\,\,\sigma_{\rm res}\,\,  \frac{\Gamma_{\rm
  res}}{2} \,\,
\delta(\varepsilon' - \varepsilon'_{\rm res})\, ,
\label{sigma}
\end{equation}
where $\sigma_{\rm res}$ is the resonance peak, $\Gamma_{\rm res}$ its
width, and $\varepsilon'_{\rm res}$ the pole in the rest frame of the
nucleus.  The factor of $1/2$ is introduced to match the integral
(i.e. total cross section) of the Breit-Wigner and the delta
function~\cite{Anchordoqui:2006pe}.

The mean interaction time  can now be readily
obtained  substituting Eq.~(\ref{sigma}) into Eq.~(\ref{app:eq:interaction}),
\begin{eqnarray}
  \frac{1}{\tau_{\rm int}^{A\gamma} (E)} & \approx & \frac{c\, \pi\,
    \sigma_{\rm res}
    \,\varepsilon'_{\rm res}\,
\Gamma_{\rm res}}{4\,
    \gamma^2}
  \int_0^\infty \frac{d \varepsilon}{\varepsilon^2}\,\,\, n(\varepsilon) \,\,\,
  \Theta (2 \gamma \varepsilon - \varepsilon'_{\rm res}) \nonumber \\
  & = & \frac{c \, \pi \, \sigma_{\rm res} \,\varepsilon'_{\rm res}\,
    \Gamma_{\rm res}}{4 \gamma^2}
  \int_{\epsilon'_{\rm res}/2 \gamma}^\infty \frac{d\varepsilon}{\varepsilon^2}\,\,
  n (\varepsilon)  \, .
 \label{A1}
\end{eqnarray}
Substituting (\ref{app:eq:photonfield}) into (\ref{A1}) we finally find~\cite{Unger:2015laa}  
\begin{widetext}
\begin{equation}
\frac{1}{\tau_{\rm int}^{A\gamma} (E)} = \frac{1}{\tau_b}
\left\{\begin{array}{ll}  \,
(E_b / E)^{\beta +1} & ~ E \leq E_b  \\
(1-\beta)/(1-\alpha) \left[\left( E_b/E \right)^{\alpha +1} -
  \left(E_b/E\right)^2 \right] +
\left(E_b/E\right)^2 & ~ E > E_b
\end{array} \right. \, ,
\end{equation}
\end{widetext}
where
\begin{equation}
\tau_b = \frac{ 2 \ E_b \ (1-\beta)} {c \, \pi \
  \sigma_{\rm res} \, A \, m_p \ \Gamma_{\rm res}
   \ n_0} \quad {\rm and} \quad
E_b = \frac{\varepsilon'_{\rm res} \ A \ m_p}{2 \varepsilon_0} .
\end{equation}
The parameters characterizing the photodisintegration cross section are:
$\sigma_{\rm res} \approx 1.45\times 10^{-27}~{\rm cm}^2 \, A$,
$\Gamma_{\rm res} = 8~{\rm MeV}$, and $\epsilon'_{\rm res} = 42.65
A^{-0.21} \, (0.925 A^{2.433})~{\rm MeV},$ for $A > 4$ ($A\leq
4$)~\cite{Karakula:1993he}. The parameters for the photopion
production cross section are: $\sigma_{\rm res} \simeq 5.0 \times
10^{-28}~{\rm cm}^2 \, A$, $\Gamma_{\rm res} = 150~{\rm MeV}$, and
$\varepsilon'_{\rm res} = (m_\Delta^2 - m_p^2)/(2 m_p) \simeq 340~{\rm
  MeV}$~\cite{Olive:2016xmw}.

In Figs.~\ref{fig:ngc2768}, \ref{fig:ngc2832}, \ref{fig:ngc3610},
\ref{fig:ngc3613}, \ref{fig:ngc4125}, and \ref{fig:ngc5128} we compare the characteristic
time scales of acceleration and energy losses for the various
representative sources given in Table~\ref{tabla} and Cen A. We consider
$^{56}$Fe and $^{28}$Si as fiducial nuclear species. In general,  photonuclear interactions  dominate
the energy losses.  By comparing the results for $\beta = -5$ and
$\beta = -4$, we can see that there is almost no dependence on the
photon spectral index. For $E \agt 10^{11}~{\rm GeV}$, the accelerated
nuclei reach the photopion production threshold, and the
significance of this process becomes more important with increasing
energy.  The photopion production threshold corresponds to an
energy-per-nucleon of $E/A \sim 10^{10}~{\rm GeV}$. Note that the
onset of photopion production is above the required
maximum \mbox{energy $= 10^{9.5}~{\rm GeV}$} for the fiducial model
in~\cite{Unger:2015laa}. Therefore, we conclude that some quasar
remnants are capable of launching heavy (and medium mass) nuclei up to
the observed maximum energies.

Though Figs.~\ref{fig:ngc2768}, \ref{fig:ngc2832}, \ref{fig:ngc3610},
\ref{fig:ngc3613},  \ref{fig:ngc4125}, and \ref{fig:ngc5128}  provide only a na\"{\i}ve illustration of
particle acceleration with energy dissipation in the black hole dynamo
it is evident that it possible to classify the one-shot acceleration
mechanism in quasar remnants according to how source parameters (mass
of the central engine and the ambient photon backgrounds) impact the
photonuclear interaction. In the first type nuclei are completely
photodisintegrated before escaping the acceleration region producing a
flux of secondary nucleons, e.g. NGC 2768, 3610, 3613,  4125, and 5128. In the second type photopion production is
the major energy damping mechanism and nuclei are able to escape
without suffering significant spallation, e.g. NGC 2832 (which is near
the center of the TA hot-spot). Combining the ideas put
forward in~\cite{Aloisio:2013hya,Unger:2015laa} we can now formulate a
new hybrid model to explain the spectral shape of extragalactic cosmic
rays, including the critical region of the ankle. Namely, the
secondary nucleons produced in the photodisintegration process would
have a soft spectral index at the source, and consequently can explain
the energy region below the ankle.  Note that after the
photodisintegration process the secondary protons can still be
accelerated to reach energies near the ankle.  On the other hand, heavy and
medium mass nuclei are emitted with a harder spectral index, and
therefore (upon propagation to Earth) can populate the spectrum above
the ankle, all the way to the GZK cutoff.  We estimate that about 16\%
of the SMBH in the universe could accelerate UHECR nuclei via the
BZ. At this stage, it is worthwhile to point out  that for SMBHs associated with jets terminating in lobes (which can be detected in
radio) like Cen A other cosmic ray acceleration mechanisms may be at play~\cite{Biermann:1987ep,Rachen:1992pg,Rieger:2011rh}.

It is also possible that the emitted nuclei could suffer additional
photodisintegration while diffusing in the source environment, as
described in~\cite{Unger:2015laa}.  Photodisintegration
  of high-energy nuclei, $A + \gamma \to A^{\prime *} +X$, is followed
  by immediate photoemission from the excited daughter nuclei,
  \mbox{$A^{\prime *} \to A' + \gamma$~\cite{Anchordoqui:2006pd}.} The
  photodisintegration process in the source environment would then
  produce a flux of gamma-rays. A suggestion of a correlation between
  UHECRs and nearby gamma-ray emitting AGNs has been put forward
  in~\cite{Nemmen:2010bp}. Such a correlation would be consistent with
  our model. However, we should exercise caution since the results
  in~\cite{Nemmen:2010bp} are given with an {\it a posteriori}
  significance, and since the new AGN correlation analyses has not
  shown an increased in significance over
  time~\cite{PierreAuger:2014yba}.

\begin{figure}
\postscript{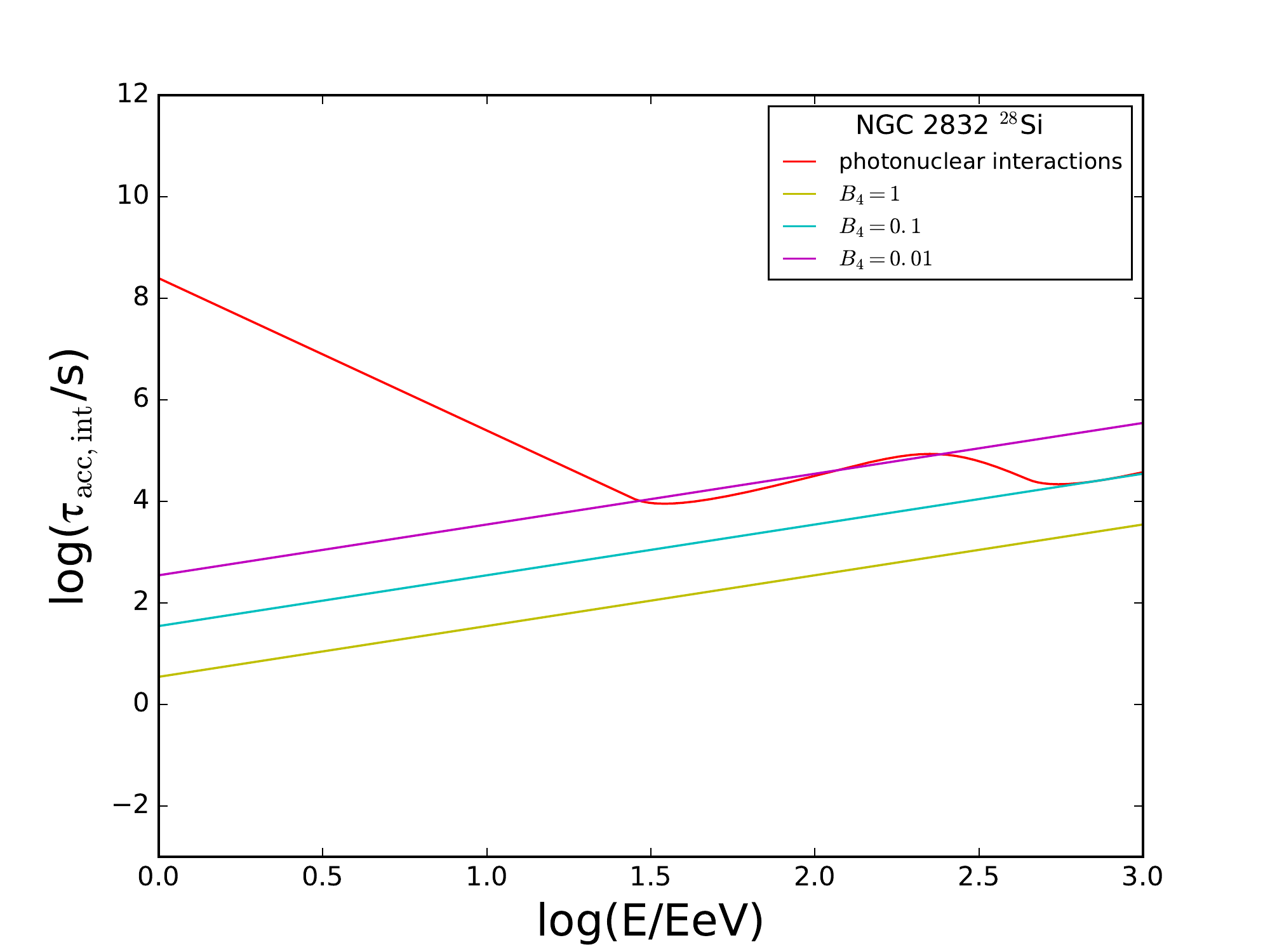}{0.99}
\caption{Comparison of acceleration and interaction time scales for
  NGC 2832. The
  bumpy curve indicates the interaction time scale for photonuclear
  interactions. The straight lines (with
  positive slope) correspond to acceleration time scales with different
magnetic field strengths; namely, $B_4 = 1,\ 0.1,\ 0.01$. We have taken $\zeta =
0.01 r_h$ and  $\beta = -4$.}
\label{fig:varyB}
\end{figure}

In closing, we study the dependence of our conclusions on the magnetic
field strength. As we have shown, the SMBHs which are capable of emitting
UHECR nuclei without suffering significant spallation are those with masses
$\agt 10^{10} M_\odot$. We can then rewrite the acceleration time
scale (\ref{muddler}) keeping $B_4$ as a variable, i.e. we take 
\begin{equation}
\tau_{\rm acc} = \frac{E}{10^{11}~{\rm GeV}}  \  \frac{1}{Z M_9 B_4}
\ \frac{r_h^2}{ \zeta c} \, .
\end{equation}
 In Fig.~\ref{fig:varyB} we
compare the acceleration and interaction time scales for different
magnetic field strengths. As we anticipated in Sec.~\ref{sec:2}, by simple inspection
of Fig.~\ref{fig:varyB} one can see that CR nuclei can be
accelerated to the highest observed energies with $B_4 \agt 0.01$.

\section{Conclusions}

\label{sec:5}

We have investigated the possibility of UHECRs being accelerated in
nearby supermassive black holes, which are dormant remnants of
previously active quasars. We have shown that these fast spinning objects
are latent dynamos that can accelerate heavy nuclei up to the maximum
observed energies. We have also shown that energy losses are dominated by
photonuclear interactions on the ambient photon fields. Armed with our
findings we postulated a classification scheme for giant elliptical
galaxies harboring supermassive black hole dynamos. In
this classification we distinguish among two unequivocal type of
cosmic ray sources according on how source
parameters (mass of the central engine and the ambient photon
backgrounds) impact the photonuclear interaction. In the first type nuclei are completely
photodisintegrated before escaping the acceleration region producing a
flux of secondary nucleons. In the
second type  photopion production is the major energy damping
mechanism and nuclei are able to escape without suffering significant spallation. In
the spirit of~\cite{Aloisio:2013hya,Unger:2015laa} we conjectured that
the spectral shape of extragalactic cosmic rays, including the pivotal
region of the ankle, can be described by a smooth superposition of
these two type of sources. Namely, the secondary nucleons produced in
the photodisintegration process would have a soft spectral index at
the source, and consequently can explain the energy region below the
ankle.  On the other hand, heavy and medium mass nuclei are emitted
with a harder spectral index, and therefore (upon propagation to
Earth) can populate the spectrum above the ankle, all the way to the
GZK cutoff.

\acknowledgments{We would like to acknowledge many useful discussions
  with our colleagues of the Pierre Auger Collaboration. We also thank
  Andrii Neronov for permission to reproduce
  Fig.~\ref{fig:field-lines}. R.J.M., R.A.C., and J.J.G. are supported
  by the U.S. National Science Foundation (NSF) Grant No.
  AST-1153335. L.A.A. is supported by NSF Grant No. PHY-1620661 and by
  the National Aeronautics and Space Administration (NASA) Grant
  No. NNX13AH52G; he thanks the Department of Physics at UNLP for its
  hospitality. Any opinions, findings, and conclusions or
  recommendations expressed in this material are those of the authors
  and do not necessarily reflect the views of the NSF or NASA.}

\onecolumngrid


\begin{thebibliography}{99}



\bibitem{Torres:2004hk} 
  D.~F.~Torres and L.~A.~Anchordoqui,
  {\color{rossoCP3} Astrophysical origins of ultrahigh energy cosmic rays},
  Rept.\ Prog.\ Phys.\  {\bf 67}, 1663 (2004)
  doi:10.1088/0034-4885/67/9/R03
  [astro-ph/0402371].





\bibitem{Kotera:2011cp} 
  K.~Kotera and A.~V.~Olinto,
  {\color{rossoCP3} The astrophysics of ultrahigh energy cosmic rays},
  Ann.\ Rev.\ Astron.\ Astrophys.\  {\bf 49}, 119 (2011)
  doi:10.1146/annurev-astro-081710-102620
  [arXiv:1101.4256 [astro-ph.HE]].


\bibitem{Abbasi:2007sv} 
  R.~U.~Abbasi {\it et al.} [HiRes Collaboration],
  {\color{rossoCP3} First observation of the Greisen-Zatsepin-Kuzmin suppression},
  Phys.\ Rev.\ Lett.\  {\bf 100}, 101101 (2008)
  doi:10.1103/PhysRevLett.100.101101
  [astro-ph/0703099].


\bibitem{Abraham:2008ru} 
  J.~Abraham {\it et al.} [Pierre Auger Collaboration],
  {\color{rossoCP3} Observation of the suppression of the flux of cosmic rays above $4\times 10^{19}$~eV},
  Phys.\ Rev.\ Lett.\  {\bf 101}, 061101 (2008)
  doi:10.1103/PhysRevLett.101.061101
  [arXiv:0806.4302 [astro-ph]].


\bibitem{Abraham:2010mj} 
  J.~Abraham {\it et al.} [Pierre Auger Collaboration],
  {\color{rossoCP3} Measurement of the energy spectrum of cosmic rays above $10^{18}$~eV using the Pierre Auger Observatory},
  Phys.\ Lett.\ B {\bf 685}, 239 (2010)
  doi:10.1016/j.physletb.2010.02.013
  [arXiv:1002.1975 [astro-ph.HE]].


\bibitem{Bird:1993yi} 
  D.~J.~Bird {\it et al.} [HiRes Collaboration],
  {\color{rossoCP3} Evidence for correlated changes in the spectrum and composition of cosmic rays at extremely high-energies},
  Phys.\ Rev.\ Lett.\  {\bf 71}, 3401 (1993)
  doi:10.1103/PhysRevLett.71.3401.

\bibitem{Allard:2005ha} D.~Allard, E.~Parizot, E.~Khan, S.~Goriely and
  A.~V.~Olinto, {\color{rossoCP3} UHE nuclei propagation and the
    interpretation of the ankle in the cosmic-ray spectrum}, Astron.\
  Astrophys.\ {\bf 443}, L29 (2005) doi:10.1051/0004-6361:200500199
  [astro-ph/0505566].


\bibitem{Allard:2005cx} 
  D.~Allard, E.~Parizot and A.~V.~Olinto,
  {\color{rossoCP3} On the transition from galactic to extragalactic cosmic-rays: spectral and composition features from two opposite scenarios},
  Astropart.\ Phys.\  {\bf 27}, 61 (2007)
  doi:10.1016/j.astropartphys.2006.09.006
  [astro-ph/0512345].



\bibitem{Greisen:1966jv} 
  K.~Greisen,
 {\color{rossoCP3} End to the cosmic ray spectrum?},
  Phys.\ Rev.\ Lett.\  {\bf 16}, 748 (1966)
  doi:10.1103/PhysRevLett.16.748.

\bibitem{Zatsepin:1966jv} 
  G.~T.~Zatsepin and V.~A.~Kuzmin,
  {\color{rossoCP3} Upper limit of the spectrum of cosmic rays},
  JETP Lett.\  {\bf 4}, 78 (1966)
  [Pisma Zh.\ Eksp.\ Teor.\ Fiz.\  {\bf 4}, 114 (1966)].




\bibitem{Hillas:1967} M. Hillas,
  {\color{rossoCP3} The energy spectrum of cosmic rays in an evolving universe},
  Phys. Lett. A {\bf 24}, 677 (1967)
doi:10.1016/0375-9601(67)91023-7.



\bibitem{Berezinsky:2002nc} 
  V.~Berezinsky, A.~Z.~Gazizov and S.~I.~Grigorieva,
  {\color{rossoCP3} On astrophysical solution to ultrahigh-energy cosmic rays},
  Phys.\ Rev.\ D {\bf 74}, 043005 (2006)
  doi:10.1103/PhysRevD.74.043005
  [hep-ph/0204357].










\bibitem{Kampert:2012mx} 
  K.~H.~Kampert and M.~Unger,
  {\color{rossoCP3} Measurements of the cosmic ray composition with air shower experiments},
  Astropart.\ Phys.\  {\bf 35}, 660 (2012)
  doi:10.1016/j.astropartphys.2012.02.004
  [arXiv:1201.0018 [astro-ph.HE]].


\bibitem{Aab:2014kda} 
  A.~Aab {\it et al.} [Pierre Auger Collaboration],
  {\color{rossoCP3} Depth of maximum of air-shower profiles at the Pierre Auger Observatory I: Measurements at energies above $10^{17.8}$ ~eV},
  Phys.\ Rev.\ D {\bf 90}, no. 12, 122005 (2014)
  doi:10.1103/PhysRevD.90.122005
  [arXiv:1409.4809 [astro-ph.HE]].

\bibitem{Aab:2014aea} 
  A.~Aab {\it et al.} [Pierre Auger Collaboration],
  {\color{rossoCP3} Depth of maximum of air-shower profiles at the Pierre Auger Observatory II: Composition implications},
  Phys.\ Rev.\ D {\bf 90}, no. 12, 122006 (2014)
  doi:10.1103/PhysRevD.90.122006
  [arXiv:1409.5083 [astro-ph.HE]].





\bibitem{Aab:2016htd} 
  A.~Aab {\it et al.} [Pierre Auger Collaboration],
   {\color{rossoCP3} Evidence for a mixed mass composition at the ‘ankle’ in the cosmic-ray spectrum},
  Phys.\ Lett.\ B {\bf 762}, 288 (2016)
  doi:10.1016/j.physletb.2016.09.039
  [arXiv:1609.08567 [astro-ph.HE]].



\bibitem{Aab:2016zth} 
  A.~Aab {\it et al.} [Pierre Auger Collaboration],
  {\color{rossoCP3}  Combined fit of spectrum and composition data as measured by the Pierre Auger Observatory},
  [arXiv:1612.07155 [astro-ph.HE]].




\bibitem{Auger:2012an} 
  P.~Abreu {\it et al.} [Pierre Auger Collaboration],
  {\color{rossoCP3} Large scale distribution of arrival directions of cosmic rays detected above $10^{18}$ eV at the Pierre Auger Observatory},
  Astrophys.\ J.\ Suppl.\  {\bf 203}, 34 (2012)
  doi:10.1088/0067-0049/203/2/34
  [arXiv:1210.3736 [astro-ph.HE]].


\bibitem{ThePierreAuger:2014nja} 
  A.~Aab {\it et al.} [Pierre Auger Collaboration],
  {\color{rossoCP3} Large scale distribution of ultrahigh energy cosmic rays detected at the Pierre Auger Observatory with zenith angles up to $80^\circ$},
  Astrophys.\ J.\  {\bf 802}, no. 2, 111 (2015)
  doi:10.1088/0004-637X/802/2/111
  [arXiv:1411.6953 [astro-ph.HE]].






\bibitem{Abbasi:2014sfa} 
  R.~U.~Abbasi {\it et al.},
  {\color{rossoCP3} Study of ultrahigh energy cosmic ray composition using Telescope Array’s Middle Drum detector and surface array in hybrid mode},
  Astropart.\ Phys.\  {\bf 64}, 49 (2014)
  doi:10.1016/j.astropartphys.2014.11.004
  [arXiv:1408.1726 [astro-ph.HE]].





\bibitem{Abbasi:2015xga} 
  R.~Abbasi {\it et al.} [Pierre Auger and Telescope Array Collaborations],
  {\color{rossoCP3} Report of the working group on the composition of ultrahigh energy cosmic rays},
  arXiv:1503.07540 [astro-ph.HE].

\bibitem{Abbasi:2014lda} 
  R.~U.~Abbasi {\it et al.} [Telescope Array Collaboration],
 {\color{rossoCP3} Indications of intermediate-scale anisotropy of cosmic rays with energy greater than 57 EeV in the Northern sky measured with the surface detector of the Telescope Array experiment},
  Astrophys.\ J.\  {\bf 790}, L21 (2014)
  doi:10.1088/2041-8205/790/2/L21
  [arXiv:1404.5890 [astro-ph.HE]].


\bibitem{Aloisio:2013hya}
  R.~Aloisio, V.~Berezinsky and P.~Blasi,
  {\color{rossoCP3} Ultrahigh energy cosmic rays: implications of Auger data for source spectra and chemical composition},
  JCAP {\bf 1410}, no. 10, 020 (2014)
  doi:10.1088/1475-7516/2014/10/020
  [arXiv:1312.7459 [astro-ph.HE]].


\bibitem{Unger:2015laa} 
  M.~Unger, G.~R.~Farrar and L.~A.~Anchordoqui,
 {\color{rossoCP3} Origin of the ankle in the ultrahigh energy cosmic ray spectrum, and of the extragalactic protons below it},
  Phys.\ Rev.\ D {\bf 92}, no. 12, 123001 (2015)
  doi:10.1103/PhysRevD.92.123001
  [arXiv:1505.02153 [astro-ph.HE]].


\bibitem{Farrar:2015ikt} 
  G.~R.~Farrar, M.~Unger and L.~A.~Anchordoqui,
  {\color{rossoCP3} Origin of the ankle in the ultra-high energy cosmic ray spectrum and  the extragalactic protons below it},
  arXiv:1512.00484 [astro-ph.HE].

\bibitem{Globus:2015xga} 
  N.~Globus, D.~Allard and E.~Parizot,
  {\color{rossoCP3} A complete model of the cosmic ray spectrum and composition across the Galactic to extragalactic transition},
  Phys.\ Rev.\ D {\bf 92}, no. 2, 021302 (2015)
  doi:10.1103/PhysRevD.92.021302
  [arXiv:1505.01377 [astro-ph.HE]].


\bibitem{Anchordoqui:2014pca} 
  L.~A.~Anchordoqui,
   {\color{rossoCP3} Neutron $\beta$-decay as the origin of IceCube’s PeV (anti)neutrinos},
  Phys.\ Rev.\ D {\bf 91}, 027301 (2015)
  doi:10.1103/PhysRevD.91.027301
  [arXiv:1411.6457 [astro-ph.HE]].

\bibitem{Taylor:2015rla} 
  A.~M.~Taylor, M.~Ahlers and D.~Hooper,
  {\color{rossoCP3} Indications of negative evolution for the sources of the highest energy cosmic rays},
  Phys.\ Rev.\ D {\bf 92}, no. 6, 063011 (2015)
  doi:10.1103/PhysRevD.92.063011
  [arXiv:1505.06090 [astro-ph.HE]].


\bibitem{Ackermann:2014usa} 
  M.~Ackermann {\it et al.} [Fermi-LAT Collaboration],
   {\color{rossoCP3} The spectrum of isotropic diffuse gamma-ray emission between 100~MeV and 820~GeV},
  Astrophys.\ J.\  {\bf 799}, 86 (2015)
  doi:10.1088/0004-637X/799/1/86
  [arXiv:1410.3696 [astro-ph.HE]].


\bibitem{TheFermi-LAT:2015ykq} 
  M.~Ackermann {\it et al.} [Fermi-LAT Collaboration],
   {\color{rossoCP3} Resolving the extragalactic $\gamma$-ray background above 50~GeV with the Fermi Large Area Telescope},
  Phys.\ Rev.\ Lett.\  {\bf 116}, no. 15, 151105 (2016)
  doi:10.1103/PhysRevLett.116.151105
  [arXiv:1511.00693 [astro-ph.CO]].



\bibitem{Liu:2016brs} 
  R.~Y.~Liu, A.~M.~Taylor, X.~Y.~Wang and F.~A.~Aharonian,
    {\color{rossoCP3} Indication of a local fog of subankle ultrahigh energy cosmic rays},
  Phys.\ Rev.\ D {\bf 94}, no. 4, 043008 (2016)
  doi:10.1103/PhysRevD.94.043008
  [arXiv:1603.03223 [astro-ph.HE]].









\bibitem{Boldt:1999ge} 
  E.~Boldt and P.~Ghosh,
  {\color{rossoCP3} Cosmic rays from remnants of quasars?},
  Mon.\ Not.\ Roy.\ Astron.\ Soc.\  {\bf 307}, 491 (1999)
  doi:10.1046/j.1365-8711.1999.02600.x
  [astro-ph/9902342].


\bibitem{Boldt:2000dx} 
  E.~Boldt and M.~Loewenstein,
  {\color{rossoCP3} Cosmic ray generation by quasar remnants: Constraints and implications},
  Mon.\ Not.\ Roy.\ Astron.\ Soc.\  {\bf 316}, L29 (2000)
  doi:10.1046/j.1365-8711.2000.03768.x
  [astro-ph/0006221].



\bibitem{Torres:2002bb} 
  D.~F.~Torres, E.~Boldt, T.~Hamilton and M.~Loewenstein,
  {\color{rossoCP3} Nearby quasar remnants and ultrahigh-energy cosmic rays},
  Phys.\ Rev.\ D {\bf 66}, 023001 (2002)
  doi:10.1103/PhysRevD.66.023001
  [astro-ph/0204419].


\bibitem{Neronov:2007mh} 
  A.~Y.~Neronov, D.~V.~Semikoz and I.~I.~Tkachev,
 {\color{rossoCP3} Ultra-high energy cosmic ray production in the polar cap regions of black hole magnetospheres},
  New J.\ Phys.\  {\bf 11}, 065015 (2009)
  doi:10.1088/1367-2630/11/6/065015
  [arXiv:0712.1737 [astro-ph]].


\bibitem{Blandford:1977ds} 
  R.~D.~Blandford and R.~L.~Znajek,
   {\color{rossoCP3} Electromagnetic extractions of energy from Kerr black holes},
  Mon.\ Not.\ Roy.\ Astron.\ Soc.\  {\bf 179}, 433 (1977).


\bibitem{Znajek} R. L. Znajek
 {\color{rossoCP3} The electric and magnetic conductivity of a Kerr  hole},
 Mon.\ Not.\ Roy.\ Astron.\ Soc.\  {\bf 185}, 833 (1978).


\bibitem{Hazard} 
C. Hazard, M. B. Mackey and A. J. Shimmins,
{\color{rossoCP3} Investigation of the radio source 3C 273 by the method of lunar occultations},
Nature {\bf 197},  1037 (1963).
doi:10.1038/1971037a0

\bibitem{Schmidt} M. Schmidt {\color{rossoCP3} 3C 273 : A star-like
  object with large red-shift},
Nature {\bf 197}, 1040 (1963).
doi:10.1038/1971040a0

\bibitem{Smith} 
H. J. Smith and D. Hoffleit
{\color{rossoCP3} Light variations in the superluminous radio galaxy 3C273},
Nature {\bf 198},  650 (1963).
doi:10.1038/198650a0


\bibitem{Zeldovich}
Ya B. Zeldovich and I. D. Novikov, 
  {\color{rossoCP3} Mass of quasi-stellar objects}, 
Doklady Acad. Nauk SSSR {\bf 158}, 811 [Sov. Phys.-Doklady {\bf 9}, 834 (1965)].


\bibitem{Salpeter:1964kb} 
  E.~E.~Salpeter,
 {\color{rossoCP3} Accretion of interstellar matter by massive objects},
  Astrophys.\ J.\  {\bf 140}, 796 (1964).
  doi:10.1086/147973



\bibitem{LyndenBell:1969yx} 
  D.~Lynden-Bell,
 {\color{rossoCP3}  Galactic nuclei as collapsed old quasars},
  Nature {\bf 223}, 690 (1969).
  doi:10.1038/223690a0


\bibitem{Antonucci:1993sg} 
  R.~Antonucci,
  {\color{rossoCP3} Unified models for active galactic nuclei and quasars},
  Ann.\ Rev.\ Astron.\ Astrophys.\  {\bf 31}, 473 (1993).
  doi:10.1146/annurev.aa.31.090193.002353



\bibitem{Urry:1995mg} 
  C.~M.~Urry and P.~Padovani,
   {\color{rossoCP3} Unified schemes for radio-loud active galactic nuclei},
  Publ.\ Astron.\ Soc.\ Pac.\  {\bf 107}, 803 (1995)
  doi:10.1086/133630
  [astro-ph/9506063].



\bibitem{Eddington} 
A. S. Eddington,
 {\color{rossoCP3}  A limiting case in the theory of radiative
   equilibrium},
 Mon.\ Not.\ Roy.\ Astron.\ Soc.\  {\bf 85}, 408 (1925)
  doi:10.1093/mnras/85.5.408

\bibitem{Shakura:1972te} 
  N.~I.~Shakura and R.~A.~Sunyaev,
     {\color{rossoCP3} Black holes in binary systems: Observational appearance},
  Astron.\ Astrophys.\  {\bf 24}, 337 (1973).





\bibitem{Rees} 
M. J. Rees,
  {\color{rossoCP3} ``Dead quasars'' in nearby galaxies},
Science {\bf 247}, 817 (1990)
doi: 10.1126/science.247.4944.817




\bibitem{Richstone:1998ky} 
  D.~Richstone {\it et al.},
    {\color{rossoCP3} Supermassive black holes and the evolution of galaxies},
  Nature {\bf 395}, A14 (1998)
  [astro-ph/9810378].


\bibitem{Ho:1998hz} 
  L.~C.~Ho,
    {\color{rossoCP3} What powers the compact radio emission in nearby elliptical and SO galaxies?},
  Astrophys.\ J.\  {\bf 510}, 631 (1999)
  doi:10.1086/306597
  [astro-ph/9808123].



\bibitem{Kormendy:2004tc} 
  J.~Kormendy and R.~Kennicutt, Jr.,
     {\color{rossoCP3} Secular evolution and the formation of pseudobulges in disk galaxies},
  Ann.\ Rev.\ Astron.\ Astrophys.\  {\bf 42}, 603 (2004)
  doi:10.1146/annurev.astro.42.053102.134024
  [astro-ph/0407343].


\bibitem{Kormendy:1995er}
  J.~Kormendy and D.~Richstone,
  {\color{rossoCP3} Inward bound: The search for supermassive black holes in galactic nuclei},
  Ann.\ Rev.\ Astron.\ Astrophys.\  {\bf 33} (1995) 581.
  doi:10.1146/annurev.aa.33.090195.003053


\bibitem{Magorrian:1997hw} 
  J.~Magorrian {\it et al.},
 {\color{rossoCP3} The demography of massive dark objects in galaxy centers},
  Astron.\ J.\  {\bf 115}, 2285 (1998)
  doi:10.1086/300353
  [astro-ph/9708072].


\bibitem{Gebhardt:2000fk} 
  K.~Gebhardt {\it et al.},
  {\color{rossoCP3} A relationship between nuclear black hole mass and galaxy velocity dispersion},
  Astrophys.\ J.\  {\bf 539}, L13 (2000)
  doi:10.1086/312840
  [astro-ph/0006289].



\bibitem{Ferrarese:2000se} 
  L.~Ferrarese and D.~Merritt,
  {\color{rossoCP3} A fundamental relation between supermassive black holes and their host galaxies},
  Astrophys.\ J.\  {\bf 539}, L9 (2000)
  doi:10.1086/312838
  [astro-ph/0006053].



\bibitem{Haring:2004hr} 
  N.~Haring and H.~W.~Rix,
  {\color{rossoCP3} On the black hole mass-bulge mass relation},
  Astrophys.\ J.\  {\bf 604}, L89 (2004)
  doi:10.1086/383567
  [astro-ph/0402376].






\bibitem{Vika:2009ef} 
  M.~Vika, S.~P.~Driver, A.~W.~Graham and J.~Liske,
    {\color{rossoCP3} The millennium galaxy catalogue: The $M_{\rm BH}
      - L_{\rm spheroid}$ derived supermassive black hole mass function},
  Mon.\ Not.\ Roy.\ Astron.\ Soc.\  {\bf 400}, 1451 (2009)
  doi:10.1111/j.1365-2966.2009.15544.x
  [arXiv:0908.2102 [astro-ph.CO]].




\bibitem{McLure:2003em} 
  R.~J.~McLure and J.~S.~Dunlop,
   {\color{rossoCP3} The cosmological evolution of quasar black-hole masses},
  Mon.\ Not.\ Roy.\ Astron.\ Soc.\  {\bf 352}, 1390 (2004)
  doi:10.1111/j.1365-2966.2004.08034.x
  [astro-ph/0310267].


\bibitem{Graham:2007qb} 
  A.~W.~Graham, S.~P.~Driver, P.~D.~Allen and J.~Liske,
    {\color{rossoCP3} The millennium galaxy catalogue: The local
      supermassive black hole mass function in early-  and late-type galaxies},
  Mon.\ Not.\ Roy.\ Astron.\ Soc.\  {\bf 378}, 198 (2007)
  doi:10.1111/j.1365-2966.2007.11770.x
  [arXiv:0704.0316 [astro-ph]].



\bibitem{Marconi:2003tg} 
  A.~Marconi, G.~Risaliti, R.~Gilli, L.~K.~Hunt, R.~Maiolino and M.~Salvati,
   {\color{rossoCP3} Local supermassive black holes, relics of active galactic nuclei and the X-ray background},
  Mon.\ Not.\ Roy.\ Astron.\ Soc.\  {\bf 351}, 169 (2004)
  doi:10.1111/j.1365-2966.2004.07765.x
  [astro-ph/0311619].


\bibitem{Tamura:2005kc} 
  N.~Tamura, K.~Ohta and Y.~Ueda,
   {\color{rossoCP3} Supermassive black hole mass functions at intermediate redshifts from spheroid and AGN luminosity functions},
  Mon.\ Not.\ Roy.\ Astron.\ Soc.\  {\bf 365}, 134 (2006)
  doi:10.1111/j.1365-2966.2005.09677.x
  [astro-ph/0509912].



\bibitem{Soltan:1982vf} 
  A.~Soltan,
 {\color{rossoCP3}  Masses of quasars},
  Mon.\ Not.\ Roy.\ Astron.\ Soc.\  {\bf 200}, 115 (1982).



\bibitem{Chokshi}
A. Chokshi and E. L. Turner
{\color{rossoCP3} Remnants of the quasars},
 Mon.\ Not.\ Roy.\ Astron.\ Soc.\  {\bf 259}, 421 (1992).
doi: 10.1093/mnras/259.3.421


\bibitem{Small92}
T. A. Small and R. D. Blandford, 
{\color{rossoCP3}  Quasar evolution and the growth of black holes},
Mon. Not. Roy. Astron. Soc. 259, 725 (1992).



\bibitem{Cao:2008pd} 
  X.~Cao and F.~Li,
  {\color{rossoCP3}  Rapidly spinning massive black holes in active galactic nuclei: evidence from the black hole mass function},
  Mon.\ Not.\ Roy.\ Astron.\ Soc.\  {\bf 390}, 561 (2008)
  doi:10.1111/j.1365-2966.2008.13800.x
  [arXiv:0808.0759 [astro-ph]].


\bibitem{Li:2011ik} 
  Y.~R.~Li, L.~C.~Ho and J.~M.~Wang,
   {\color{rossoCP3} Cosmological evolution of supermassive black
     holes 
I: mass function at 0},  
Astrophys.\ J.\  {\bf 742}, 33 (2011)
  doi:10.1088/0004-637X/742/1/33
  [arXiv:1109.0089 [astro-ph.CO]].




\bibitem{Shankar:2007zg} 
  F.~Shankar, D.~H.~Weinberg and J.~Miralda-Escude,
  {\color{rossoCP3}  Self-consistent models of the AGN and black hole populations: Duty cycles, accretion rates, and the mean radiative efficiency},
  Astrophys.\ J.\  {\bf 690}, 20 (2009)
  doi:10.1088/0004-637X/690/1/20
  [arXiv:0710.4488 [astro-ph]].

\bibitem{Yu:2002sq} 
  Q.~j.~Yu and S.~Tremaine,
 {\color{rossoCP3}   Observational constraints on growth of massive black holes},
  Mon.\ Not.\ Roy.\ Astron.\ Soc.\  {\bf 335}, 965 (2002)
  doi:10.1046/j.1365-8711.2002.05532.x
  [astro-ph/0203082].



\bibitem{Volonteri:2007tu} 
  M.~Volonteri, M.~Sikora and J.~P.~Lasota,
   {\color{rossoCP3} Black-hole spin and galactic morphology},
  Astrophys.\ J.\  {\bf 667}, 704 (2007)
  doi:10.1086/521186
  [arXiv:0706.3900 [astro-ph]].


\bibitem{Anchordoqui:2016dcp} 
  L.~A.~Anchordoqui,
   {\color{rossoCP3} Neutrino lighthouse powered by Sagittarius A$^*$ disk dynamo},
  Phys.\ Rev.\ D {\bf 94}, 023010 (2016)
  doi:10.1103/PhysRevD.94.023010
  [arXiv:1606.01816 [astro-ph.HE]].

\bibitem{Blandford:1978}
R. D. Blandford and M. J.  Rees, 
 {\color{rossoCP3} Extended and compact extragalactic interpretation and theory}
Phys.\ Scripta {\bf  17}, 265 (1978).


\bibitem{Marcha} M. J. M. Marcha, I. W. A. Browne, C. D. Impey and
  P. S. Smith,
 {\color{rossoCP3} Optical spectroscopy and polarization of a new sample of optically bright flat radio spectrum sources}
Mon.\ Not.\ Roy.\ Astron.\ Soc.\  {\bf 281}, 425 (1996)
 doi:10.1093/mnras/281.2.425.

\bibitem{Giommi:2011sn} 
  P.~Giommi, P.~Padovani, G.~Polenta, S.~Turriziani, V.~D'Elia and S.~Piranomonte,
   {\color{rossoCP3} A simplified view of blazars: clearing the fog around long-standing selection effects},
  Mon.\ Not.\ Roy.\ Astron.\ Soc.\  {\bf 420}, 2899 (2012)
  doi:10.1111/j.1365-2966.2011.20044.x
  [arXiv:1110.4706 [astro-ph.CO]].


\bibitem{Cavaliere:2001gb} 
  A.~Cavaliere and V.~D'Elia,
    {\color{rossoCP3}  The blazar main sequence},
  Astrophys.\ J.\  {\bf 571}, 226 (2002)
  doi:10.1086/339778
  [astro-ph/0106512].


\bibitem{Boettcher:2001jp} 
  M.~Boettcher and C.~D.~Dermer,
    {\color{rossoCP3}  An evolutionary scenario for blazar unification},
  Astrophys.\ J.\  {\bf 564}, 86 (2002)
  doi:10.1086/324134
  [astro-ph/0106395].


\bibitem{Ajello:2011zi} 
  M.~Ajello {\it et al.},
    {\color{rossoCP3}  The luminosity function of Fermi-detected flat-spectrum radio quasars},
  Astrophys.\ J.\  {\bf 751}, 108 (2012)
  doi:10.1088/0004-637X/751/2/108
  [arXiv:1110.3787 [astro-ph.CO]].



\bibitem{Ajello:2013lka} 
  M.~Ajello {\it et al.},
    {\color{rossoCP3}  The cosmic evolution of Fermi BL Lacertae objects},
  Astrophys.\ J.\  {\bf 780}, 73 (2014)
  doi:10.1088/0004-637X/780/1/73
  [arXiv:1310.0006 [astro-ph.CO]].



\bibitem{Maraschi:2002pp} 
  L.~Maraschi and F.~Tavecchio,
   {\color{rossoCP3}  The jet-disk connection and blazar unification},
  Astrophys.\ J.\  {\bf 593}, 667 (2003)
  doi:10.1086/342118
  [astro-ph/0205252].



\bibitem{Lintott:2009wq} 
  C.~Lintott {\it et al.},
   {\color{rossoCP3}  Galaxy zoo : ``Hanny's Voorwerp'', a quasar light echo?},
  Mon.\ Not.\ Roy.\ Astron.\ Soc.\  {\bf 399}, 129 (2009)
  doi:10.1111/j.1365-2966.2009.15299.x
  [arXiv:0906.5304 [astro-ph.CO]].


\bibitem{Jozsa:2009qm} 
  G.~I.~G.~Jozsa {\it et al.},
    {\color{rossoCP3}  Revealing Hanny's Voorwerp: radio observations of IC 2497},
  Astron.\ Astrophys.\  {\bf 500}, L33 (2009)
  doi:10.1051/0004-6361/200912402
  [arXiv:0905.1851 [astro-ph.CO]].


\bibitem{Rampadarath:2010yv} 
  H.~Rampadarath {\it et al.},
    {\color{rossoCP3} Hanny's Voorwerp: Evidence of AGN activity and a nuclear starburst in the central regions of IC 2497},
  Astron.\ Astrophys.\  {\bf 517}, L8 (2010)
  doi:10.1051/0004-6361/201014782
  [arXiv:1006.4096 [astro-ph.GA]].

\bibitem{Schawinski:2010up} 
  K.~Schawinski {\it et al.},
   {\color{rossoCP3}  The sudden death of the nearest quasar},
  Astrophys.\ J.\  {\bf 724}, L30 (2010)
  doi:10.1088/2041-8205/724/1/L30
  [arXiv:1011.0427 [astro-ph.CO]].

\bibitem{PierreAuger:2014yba} 
  A.~Aab {\it et al.} [Pierre Auger Collaboration],
     {\color{rossoCP3}  Searches for anisotropies in the arrival directions of the highest energy cosmic rays detected by the Pierre Auger Observatory},
  Astrophys.\ J.\  {\bf 804}, no. 1, 15 (2015)
  doi:10.1088/0004-637X/804/1/15
  [arXiv:1411.6111 [astro-ph.HE]].


\bibitem{Abreu:2010ab} 
  P.~Abreu {\it et al.} [Pierre Auger Collaboration],
     {\color{rossoCP3}  Update on the correlation of the highest energy cosmic rays with nearby extragalactic matter},
  Astropart.\ Phys.\  {\bf 34}, 314 (2010)
  doi:10.1016/j.astropartphys.2010.08.010
  [arXiv:1009.1855 [astro-ph.HE]].


\bibitem{Abraham:2007si} 
  J.~Abraham {\it et al.} [Pierre Auger Collaboration],
     {\color{rossoCP3}  Correlation of the highest-energy cosmic rays with the positions of nearby active galactic nuclei},
  Astropart.\ Phys.\  {\bf 29}, 188 (2008)
  Erratum: [Astropart.\ Phys.\  {\bf 30}, 45 (2008)]
  doi:10.1016/j.astropartphys.2008.06.004, 10.1016/j.astropartphys.2008.01.002
  [arXiv:0712.2843 [astro-ph]].


\bibitem{Abraham:2007bb} 
  J.~Abraham {\it et al.} [Pierre Auger Collaboration],
     {\color{rossoCP3}  Correlation of the highest energy cosmic rays with nearby extragalactic objects},
  Science {\bf 318}, 938 (2007)
  doi:10.1126/science.1151124
  [arXiv:0711.2256 [astro-ph]].




\bibitem{Isola:2003jk} 
  C.~Isola, G.~Sigl and G.~Bertone,
    {\color{rossoCP3}  Ultrahigh energy cosmic rays from quasar remnants},
  astro-ph/0312374.

\bibitem{Cavallo:1978}
G. Cavallo,
  {\color{rossoCP3} On the sources of ultra-high energy cosmic rays},
Astron.\ Astrophys.\ {\bf 65}, 415 (1978).

\bibitem{Romero:1995tn} 
  G.~E.~Romero, J.~A.~Combi, L.~A.~Anchordoqui and S.~E.~Perez Bergliaffa,
    {\color{rossoCP3} A possible source of extragalactic cosmic rays with arrival energies beyond the GZK cutoff},
  Astropart.\ Phys.\  {\bf 5}, 279 (1996)
  doi:10.1016/0927-6505(96)00029-1
  [gr-qc/9511031].


\bibitem{Dawson:2013wsa} 
  B.~R.~Dawson {\it et al.} [Pierre Auger and Yakutsk and Telescope Array Collaborations],
   {\color{rossoCP3}  The energy spectrum of cosmic rays at the highest energies},
  EPJ Web Conf.\  {\bf 53}, 01005 (2013)
  doi:10.1051/epjconf/20135301005
  [arXiv:1306.6138 [astro-ph.HE]].


\bibitem{Adams:2017fjh} 
  J.~H.~Adams {\it et al.},
  {\color{rossoCP3}  White paper on EUSO-SPB2},
  arXiv:1703.04513 [astro-ph.HE].











\bibitem{Kerr:1963ud} 
  R.~P.~Kerr,
    {\color{rossoCP3}  Gravitational field of a spinning mass as an example of algebraically special metrics},
  Phys.\ Rev.\ Lett.\  {\bf 11}, 237 (1963).
  doi:10.1103/PhysRevLett.11.237


\bibitem{Schwarzschild:1916uq} 
  K.~Schwarzschild,
  {\color{rossoCP3}  On the gravitational field of a mass point according to Einstein's theory},
  Sitzungsber.\ Preuss.\ Akad.\ Wiss.\ Berlin (Math.\ Phys.\ ) {\bf 1916}, 189 (1916)
  [physics/9905030].



\bibitem{Thorne:1974ve} 
  K.~S.~Thorne,
{\color{rossoCP3}  Disk accretion onto a black hole 2: Evolution of the hole},
  Astrophys.\ J.\  {\bf 191}, 507 (1974).
  doi:10.1086/152991



\bibitem{Visser:2007fj} 
  M.~Visser,
     {\color{rossoCP3} The Kerr spacetime: A Brief introduction},
  arXiv:0706.0622 [gr-qc].


\bibitem{Rieger:2011ch} 
  F.~M.~Rieger,
    {\color{rossoCP3} Non-thermal processes in black-hole-jet magnetospheres},
  Int.\ J.\ Mod.\ Phys.\ D {\bf 20}, 1547 (2011)
  doi:10.1142/S0218271811019712
  [arXiv:1107.2119 [astro-ph.CO]].


\bibitem{Penrose:1969pc} 
  R.~Penrose,
    {\color{rossoCP3} Gravitational collapse: The role of general relativity},
  Riv.\ Nuovo Cim.\  {\bf 1}, 252 (1969)
  [Gen.\ Rel.\ Grav.\  {\bf 34}, 1141 (2002)].

\bibitem{Penrose:1971uk} 
  R.~Penrose and R.~M.~Floyd,
  {\color{rossoCP3}   Extraction of rotational energy from a black hole},
  Nature {\bf 229}, 177 (1971).


\bibitem{Wald:1974np} 
  R.~M.~Wald,
   {\color{rossoCP3} Black hole in a uniform magnetic field},
  Phys.\ Rev.\ D {\bf 10}, 1680 (1974).
  doi:10.1103/PhysRevD.10.1680





\bibitem{Blasi:2000xm} 
  P.~Blasi, R.~I.~Epstein and A.~V.~Olinto,
  {\color{rossoCP3} Ultrahigh energy cosmic rays from young neutron star winds},
  Astrophys.\ J.\  {\bf 533}, L123 (2000)
  doi:10.1086/312626
  [astro-ph/9912240].



\bibitem{Kotera:2015pya} 
  K.~Kotera, E.~Amato and P.~Blasi,
  {\color{rossoCP3} The fate of ultrahigh energy nuclei in the immediate environment of young fast-rotating pulsars},
  JCAP {\bf 1508}, no. 08, 026 (2015)
  doi:10.1088/1475-7516/2015/08/026
  [arXiv:1503.07907 [astro-ph.HE]].








\bibitem{Ochelkov} 
Y. P. Ochelkov and V. V. Usov,
{\color{rossoCP3} Curvature radiation of relativistic particles in the
  magnetosphere of pulsars}
 Astrophys.\ Space Sci.\  {\bf 69},  439 (1980)
doi:10.1007/BF00661929.



\bibitem{Levinson:2000nx} 
  A.~Levinson,
  {\color{rossoCP3} Particle acceleration and curvature TeV emission by rotating supermassive black holes},
  Phys.\ Rev.\ Lett.\  {\bf 85}, 912 (2000).
  doi:10.1103/PhysRevLett.85.912
 [hep-ph/0002020].


\bibitem{Levinson:2002ea} 
  A.~Levinson and E.~Boldt,
  {\color{rossoCP3} UHECR production by a compact black hole dynamo: Application to Sgr A$^*$},
  Astropart.\ Phys.\  {\bf 16}, 265 (2002).
  doi:10.1016/S0927-6505(01)00116-5
 [astro-ph/0012314].


\bibitem{Whysong:2002ks} 
  D.~Whysong and R.~Antonucci,
   {\color{rossoCP3} Thermal emission as a test for hidden nuclei in nearby radio galaxies},
  Astrophys.\ J.\  {\bf 602}, 116 (2004)
  doi:10.1086/380828
  [astro-ph/0207385].


\bibitem{Meisenheimer:2007zv} 
  K.~Meisenheimer {\it et al.},
   {\color{rossoCP3} Resolving the innermost parsec of Centaurus A at mid-infrared wavelengths},
  Astron.\ Astrophys.\  {\bf 471}, 453 (2007)
  doi:10.1051/0004-6361:20066967
  [arXiv:0707.0177 [astro-ph]].


\bibitem{NED} {\tt
  https://ned.ipac.caltech.edu.}

\bibitem{Wiklind}
T. Wiklind and C. Henkel,
  {\color{rossoCP3} Cold dust in elliptical galaxies}
Astron. Astrophys. {\bf 297}, L71 (1995).


\bibitem{Marshall:1988} M. Joy, D. F. Lester, P. H. Harvey and
 H. B. Ellis,
  {\color{rossoCP3} The origing of the infrared luminosity in
    Centaurus A}
Astrophys.\ J.\ {\bf 326}, 662 (1988).

\bibitem{Eckart}
A. Eckart, M. Cameron, H. Rothermel, W. Wild, H. Zinnecker,
G. Rydbeck, M. Olberg and T. Wiklind,
  {\color{rossoCP3} Observations of CO isotopic emission and the
    far-infrared continuum of Centaurus A}
Astrophys.\ J.\ {\bf 363}, 451 (1990).


\bibitem{Wild:2000hq} 
  W.~Wild and A.~Eckart,
    {\color{rossoCP3} Dense gas in the dust lane of Centaurus A},
  Astron.\ Astrophys.\  {\bf 359}, 483 (2000)
  [astro-ph/0005401].


\bibitem{Stawarz:2006ku} 
  L.~Stawarz, F.~Aharonian, S.~Wagner and M.~Ostrowski,
    {\color{rossoCP3} Absorption of nuclear gamma-rays on the starlight radiation in FR I sources: the case of Centaurus A},
  Mon.\ Not.\ Roy.\ Astron.\ Soc.\  {\bf 371}, 1705 (2006)
  doi:10.1111/j.1365-2966.2006.10807.x
  [astro-ph/0605721].


\bibitem{Graham:1981} J. A. Graham and R. M. Price,
  {\color{rossoCP3} The gaseous filaments in the northeast halo region of NGC 5128 (Centaurus A)}
Astrophys.\ J.\ {\bf 247}, 813 (1981).



\bibitem{Crocker:2008dk} 
  A.~F.~Crocker, M.~Bureau, L.~M.~Young and F.~Combes,
 {\color{rossoCP3}   The molecular polar disc in NGC 2768},
  Mon.\ Not.\ Roy.\ Astron.\ Soc.\  {\bf 386}, 1811 (2008)
  doi:10.1111/j.1365-2966.2008.13177.x
  [arXiv:0803.0426 [astro-ph]].



\bibitem{Temi:2009ue} 
  P.~Temi, F.~Brighenti and W.~G.~Mathews,
  {\color{rossoCP3}  Spitzer observations of passive and star forming early-type galaxies: an infrared color-color sequence},
  Astrophys.\ J.\  {\bf 707}, 890 (2009)
  doi:10.1088/0004-637X/707/2/890
  [arXiv:0911.0720 [astro-ph.CO]].

\bibitem{Brown:2011xh} 
  M.~J.~I.~Brown, B.~T.~Jannuzi, D.~J.~E.~Floyd and J.~R.~Mould,
   {\color{rossoCP3} The ubiquitous radio continuum emission from the most massive early-type galaxies},
  Astrophys.\ J.\  {\bf 731}, L41 (2011)
  doi:10.1088/2041-8205/731/2/L41
  [arXiv:1103.2828 [astro-ph.CO]].



\bibitem{Temi:2003sj} 
  P.~Temi, F.~Brighenti, W.~G.~Mathews and J.~D.~Bregman,
  {\color{rossoCP3} Cold dust in early-type galaxies I: Observations},
  Astrophys.\ J.\ Suppl.\  {\bf 151}, 237 (2004)
  doi:10.1086/381963
  [astro-ph/0312248].



\bibitem{Olive:2016xmw} 
  C.~Patrignani {\it et al.} [Particle Data Group],
   {\color{rossoCP3}  Review of Particle Physics},
  Chin.\ Phys.\ C {\bf 40}, no. 10, 100001 (2016).
  doi:10.1088/1674-1137/40/10/100001



\bibitem{Marconi:2005qx} 
  A.~Marconi, G.~Pastorini, F.~Pacini, D.~J.~Axon, A.~Capetti, D.~Macchetto, A.~M.~Koekemoer and E.~J.~Schreier,
   {\color{rossoCP3}  The supermassive black hole in Centaurus A: A benchmark for gas kinematical measurements},
  Astron.\ Astrophys.\  {\bf 448}, 921 (2006)
  doi:10.1051/0004-6361:20053853
  [astro-ph/0507435].




\bibitem{Neumayer:2007ic} 
  N.~Neumayer, M.~Cappellari, J.~Reunanen, H.~W.~Rix, P.~P.~van der Werf, P.~T.~de Zeeuw and R.~I.~Davies,
   {\color{rossoCP3}  The central parsecs of Centaurus A: high excitation gas, a molecular disk, and the mass of the black hole},
  Astrophys.\ J.\  {\bf 671}, 1329 (2007)
  doi:10.1086/523039
  [arXiv:0709.1877 [astro-ph]].



\bibitem{Cappellari:2008db} 
  M.~Cappellari, N.~Neumayer, J.~Reunanen, P.~P.~van der Werf, P.~T.~de Zeeuw and H.-W.~Rix,
    {\color{rossoCP3} The mass of the black hole in Centaurus A from SINFONI AO-assisted integral-field observations of stellar kinematics},
  Mon.\ Not.\ Roy.\ Astron.\ Soc.\  {\bf 394}, 660 (2009)
  doi:10.1111/j.1365-2966.2008.14377.x
  [arXiv:0812.1000 [astro-ph]].








\bibitem{Stecker:1969fw} 
  F.~W.~Stecker,
 {\color{rossoCP3}  Photodisintegration of ultrahigh-energy cosmic rays by the universal radiation field},
  Phys.\ Rev.\  {\bf 180}, 1264 (1969).
  doi:10.1103/PhysRev.180.1264


\bibitem{Anchordoqui:2006pe} 
  L.~A.~Anchordoqui, J.~F.~Beacom, H.~Goldberg, S.~Palomares-Ruiz and T.~J.~Weiler,
  {\color{rossoCP3} TeV $\gamma$-rays and neutrinos from photodisintegration of nuclei in Cygnus OB2},
  Phys.\ Rev.\ D {\bf 75}, 063001 (2007)
  doi:10.1103/PhysRevD.75.063001
  [astro-ph/0611581].


\bibitem{Karakula:1993he} 
  S.~Karakula and W.~Tkaczyk,
  {\color{rossoCP3}  The formation of the cosmic ray energy spectrum by a photon field},
  Astropart.\ Phys.\  {\bf 1}, 229 (1993).
  doi:10.1016/0927-6505(93)90023-7


\bibitem{Biermann:1987ep} 
  P.~L.~Biermann and P.~A.~Strittmatter,
    {\color{rossoCP3} Synchrotron emission from shock waves in active galactic nuclei},
  Astrophys.\ J.\  {\bf 322}, 643 (1987).
  doi:10.1086/165759


\bibitem{Rachen:1992pg} 
  J.~P.~Rachen and P.~L.~Biermann,
    {\color{rossoCP3} Extragalactic ultrahigh-energy cosmic rays I: Contribution from hot spots in FR-II radio galaxies},
  Astron.\ Astrophys.\  {\bf 272}, 161 (1993)
  [astro-ph/9301010].




\bibitem{Rieger:2011rh} 
  F.~M.~Rieger,
  {\color{rossoCP3}  Cen A as gamma- and UHE cosmic-ray source},
  Mem.\ Soc.\ Ast.\ It.\  {\bf 83}, 127 (2012)
  [arXiv:1108.4565 [astro-ph.HE]].


\bibitem{Anchordoqui:2006pd} 
  L.~A.~Anchordoqui, J.~F.~Beacom, H.~Goldberg, S.~Palomares-Ruiz and T.~J.~Weiler,
   {\color{rossoCP3}  TeV gamma-rays from photo-disintegration/de-excitation of cosmic-ray nuclei},
  Phys.\ Rev.\ Lett.\  {\bf 98}, 121101 (2007)
  doi:10.1103/PhysRevLett.98.121101
  [astro-ph/0611580].



\bibitem{Nemmen:2010bp} 
  R.~S.~Nemmen, C.~Bonatto and T.~Storchi-Bergmann,
   {\color{rossoCP3}  A correlation between the highest energy cosmic rays and nearby active galactic nuclei detected by Fermi},
  Astrophys.\ J.\  {\bf 722}, 281 (2010)
  doi:10.1088/0004-637X/722/1/281
  [arXiv:1007.5317 [astro-ph.HE]].








\end{thebibliography}
\end{document}